\documentclass[11pt]{article}
\usepackage[english]{babel}
\usepackage{booktabs}
\usepackage{subfigure}
\usepackage{amsmath}
\usepackage{amscd,amsfonts,amstext,
	amsthm,amssymb,color,graphicx,commath,
	latexsym,mathrsfs,tikz,tkz-fct,listofsymbols,dsfont,subcaption}
\usepackage[subfigure]{tocloft}
\usepackage[round]{natbib}
\usepackage{longtable}
\usepackage{url}

\usepackage{diagbox}
\usepackage{threeparttable}
\usepackage{verbatim}

\usepackage[title]{appendix}
\allowdisplaybreaks

\usepackage{float}
\usepackage{titlesec}
\usepackage[letterpaper,left=2.5cm,top=1.54cm,bottom=2.54cm,right=2.5cm]{geometry}
\usepackage{epstopdf}
\usepackage{color}
\usepackage{multirow}
\usepackage{enumerate}
\usepackage{authblk}
\usepackage{hyperref}
\usepackage{pdfpages}
\usepackage{lineno}
\usepackage{algorithmic}
\usepackage{algorithm}

\titleformat*{\section}{\large\bfseries}

\newcommand{\rmnum}[1]{\lowercase\expandafter{\romannumeral #1\relax}}
\newcommand{\Rmnum}[1]{\uppercase\expandafter{\romannumeral #1\relax}}
\DeclareMathOperator*{\argmin}{arg\,min}

\numberwithin{equation}{section}

\newcommand{\RNum}[1]{\uppercase\expandafter{\romannumeral #1\relax}}
\newcommand{\rnum}[1]{\lowercase\expandafter{\romannumeral #1\relax}}

\usepackage{hyperref}

\definecolor{darkblue}{rgb}{0.1,0.1,0.7}
\definecolor{darkred}{rgb}{0.9,0.1,0.1}

\hypersetup{colorlinks, allcolors= darkblue}

\title{\vspace{+0.5cm}\Large\textbf{{Fairness-Aware Insurance Pricing: A Multi-Objective Optimization Approach}}\thanks{Corresponding author: Xinyue Fan. E-mail: \texttt{fxy0108@connect.hku.hk}}}

\author{Tim J. Boonen}
\author{Xinyue Fan}
\author{Zixiao Quan}

\affil{\emph{\small Department of Statistics and Actuarial Science, School of Computing and Data Science, University of Hong Kong, Hong Kong SAR, China.}}

\begin{document}

\maketitle

\begin{abstract}
Machine learning improves predictive accuracy in insurance pricing but exacerbates trade-offs between competing fairness criteria across 
different discrimination measures, challenging regulators and insurers to reconcile profitability with equitable outcomes. While existing fairness-aware models 
offer partial solutions under GLM and XGBoost estimation methods, they remain constrained by single-objective optimization, failing to holistically navigate a conflicting landscape of accuracy, group fairness, individual fairness, and counterfactual fairness. To address this, we propose a novel multi-objective optimization framework that jointly optimizes all four criteria via the Non-dominated Sorting Genetic Algorithm II (NSGA-II), generating a diverse Pareto front of trade-off solutions. We use a specific selection mechanism to extract a premium on this front. Our results show that XGBoost outperforms GLM in accuracy but amplifies fairness disparities; the Orthogonal model excels in group fairness, while Synthetic Control leads in individual and counterfactual fairness. Our method consistently achieves a balanced compromise, outperforming single-model approaches. 
\end{abstract}


\newpage

\section{Introduction}\label{intro}
The global insurance industry currently faces two competing imperatives: actuarial soundness and regulatory demands for fairness. For decades, the foundation of insurance pricing has been risk-based differentiation. This practice is economically essential for solvency and legally sanctioned for market efficiency. This foundational principle is now under intense scrutiny. Regulators worldwide are tightening restrictions on the use of policyholder data as they seek to eliminate unfair discrimination. Pivotal regulations, such as the EU’s Gender Directive, demonstrate a paradigm shift. Insurers no longer just predict risk; they must now design pricing models that are compliant, equitable, and profitable in a complex legal landscape.

The integration of machine learning (ML) presents a critical paradox. On the one hand, ML models promise unprecedented predictive power. They use vast datasets such as telematics and digital footprints to create hyper-personalized premiums, which offer a clear path to greater profitability \citep{eling2020impact}. On the other hand, this sophistication introduces profound risks. The opaque nature of many ML algorithms can inadvertently amplify discrimination. Models can infer protected attributes such as race or gender even when these features are removed \citep{barocas2016big, prince2019proxy, chibanda2022defining}. They do this through seemingly neutral proxy variables like ZIP codes or occupations, which lead to indirect discrimination \citep{tobler2008limits}. This form of bias is often unintentional and difficult to detect \citep{prince2019proxy}. It poses a significant compliance and reputational threat. Ultimately, it challenges the boundaries of acceptable underwriting.

This situation necessitates an examination of fairness in actuarial terms beyond standard actuarial fairness. The machine learning literature offers many definitions of fairness. 
These include group fairness, which seeks statistical parity, and individual fairness, which treats similar individuals similarly \citep{morse2022ends}. These concepts often conflict with the core principle of risk-based pricing. Furthermore, research has shown that many fairness metrics are mutually incompatible, which means that satisfying one metric may violate another \citep{berk2021fairness}. The field of fair ML has proposed technical solutions, but their application in insurance remains critically underexplored. A clear framework is needed to navigate the trade-offs between accuracy, actuarial justification, and competing fairness goals.

While foundational research has begun to map this territory, significant gaps remain. Initial studies establish formal frameworks for a "discrimination-free" price \citep{lindholm2022discrimination} and compare linear models \citep{frees2023discriminating}. Recent work links insurance regulations to modern ML models such as XGBoost \citep{xin2024antidiscrimination}. However, these pioneering works focus largely on a single definition of fairness. They prioritize group-level demographic parity, but the richer concepts of individual and causal fairness remain less explored. Although recent work by \citet{cote2025fair} addresses fairness in insurance pricing through causal inference and evaluates the fairness properties of five methodological approaches, the existing studies lack a systematic method for managing the inherent trade-offs  between fairness and accuracy in the pricing models.

This study provides a comprehensive framework to address these critical gaps. We make four main contributions. First, we systematically evaluate a wide range of fairness-aware pricing models. Our evaluation covers preprocessing, inprocessing, and postprocessing techniques under both GLM and XGBoost estimation methods. Second, we expand the analysis beyond group fairness. We assess models against the rigorous standards of individual and counterfactual fairness. Third, we recognize that no single model is perfect. We therefore introduce a multi-objective optimization framework. This approach uses model ensembling and the NSGA-II algorithm to identify optimal trade-offs. We then use the Technique for Order of Preference by Similarity to Ideal Solution (TOPSIS) method to select a pragmatically balanced solution, which offers a clear decision-support tool. Finally, our findings provide actionable insights for insurers and regulators to design pricing systems that are accurate, fair, and compliant. For clarity, we focus on expected loss estimation under regulatory constraints and abstract from downstream market dynamics such as competition and loyalty. 

The paper is organized as follows. Section \ref{def} summarizes key fairness definitions and Section \ref{sec:method} details the evaluation metrics and introduces our fairness-aware models. Section \ref{analysis} presents the empirical analysis and results. Section \ref{sec:conclusion} concludes with some implications for insurance regulation and practice.

\section{Fairness Criteria}\label{def}
To address the gap between fair machine learning and insurance applications mentioned in Section~\ref{intro}, this study systematically examines prominent notions of fairness in the field of machine learning and develops tailored metrics to quantify and assess fairness within actuarial pricing models in Section~\ref{sec:method}.

To formalize fairness in insurance pricing, we adopt the following notation. Let $Y$ be the actual claim amount and $\hat{Y}$ be the predicted pure premium that excludes the expenses and profit loadings. 
Let $X$ denote the non-protected variables and $D$ the sensitive attribute. We assume $D$ to be a categorical variable with only two levels $a$ and $b$.
The space $\mathcal{X}$ denotes the feature space of the non-protected variables, and $\mathcal{D}$ denotes the sensitive attribute's feature space.
Finally, $\mu$ represents the mapping from customer attributes to the estimated premium, which corresponds to the models described in Section \ref{model}.

Fairness in insurance pricing can be broadly categorized into two types based on the unit of analysis: individual fairness and group fairness. Individual fairness focuses on the level of the individual, which ensures that similar policyholders receive similar treatment. On the other hand, group fairness operates at the group level and aims for equitable outcomes across different demographic groups. These two concepts incorporate fundamentally different principles of equity. Understanding their distinctions is essential for evaluating fairness‑aware pricing models.

To situate our analysis within the broader policy context, we map prevailing insurance pricing regulations from major jurisdictions to the fairness frameworks introduced in this paper. Each law or guideline is categorized by the type of discrimination or fairness principle it targets. This mapping of statutory texts, court rulings, and regulatory interpretations is provided in Appendix~\ref{reg}.

\subsection{Group Fairness}
\subsubsection{Demographic Parity (DP)}
A predictor $\hat{Y}$ satisfies demographic parity if $$E(\hat{Y}|D=a)=E(\hat{Y}|D=b).$$
Demographic Parity requires the two groups to share a similar average premium. This criterion is also known as statistical parity and implies that the predictor $\hat{Y}$ is statistically independent of $D$. 
By enforcing equality in average premiums between groups, demographic parity disregards underlying differences in risk distributions between groups. As a result, achieving this form of fairness often entails cross-subsidization at the cost of being less competitive. 
This is a situation where lower-risk individuals (e.g., within a lower-risk group) effectively subsidize higher-risk individuals, which raises concerns about actuarial fairness and potential adverse selection. 
For example, the community rating systems implemented in Ireland’s private health insurance sector adhere strictly to this criterion by ensuring uniform premiums within plans. These premiums are independent of individual health or risk characteristics. However, market-wide fairness may be compromised if the underlying risk equalization scheme is undermined or ineffective \citep{turner2013community}.

We quantify demographic parity via the disparity impact ratio, which is a common benchmark for group fairness. This metric represents the ratio of average expected premiums across different groups. We follow the definition proposed by \citet{xin2024antidiscrimination}:
\[
\text{Disparity Impact Ratio} = \frac{E(\hat{Y}|D=a)}{E(\hat{Y}|D=b)}.\]
In our context, $D = a$ denotes the female group and $D = b$ denotes the male group. Consequently, the disparity impact ratio (DIR) is defined as the ratio of the expected predicted pure premiums for females to that for males.

\subsubsection{Conditional Demographic Parity (CDP)}\label{cdp}
A predictor $\hat{Y}$ satisfies conditional demographic parity if the outcome is equal across groups after controlling for a certain subset of non-protected variables denoted by $X_p$. This is expressed mathematically as: $$E(\hat{Y}|X_p=x,D=a)=E(\hat{Y}|X_p=x,D=b),$$
where the set $X_p\subseteq X$ represents the permitted variables that can be used. It is important to note that $X_p$ is only a subset of the non-protected variables $X$. These variables are termed “permitted” because they are allowed to be used when evaluating conditional demographic parity (see Section~\ref{cdp}). 
Non‑protected variables are those allowed in prediction, whereas the subset $X_p$ ``‘permitted variables'’) denotes those additionally allowed to appear in fairness assessments.

Conditional demographic parity requires the average premium to be equal across groups defined by a protected attribute, conditional on a specified set of non-protected attributes. It relaxes demographic parity by allowing some risk differentiation via $X_p$ and thereby sacrifices some group fairness in favor of actuarial fairness \citep{xin2024antidiscrimination}. The selection of the feature set $X_p$ depends on whether these variables are legally and ethically permitted to influence the final premium and whether they can be legitimately used in the risk classification process \citep{corbett2017algorithmic}. A permissible variable $X_p$ must capture genuine differences in expected loss or risk exposure consistently across demographic groups, and must not act as a proxy for protected attributes \citep{hardt2016equality}. The set of such admissible variables may vary depending on the type of insurance, reflecting differing actuarial justifications and regulatory standards. In the absence of any permitted conditioning variables, CDP reduces to Demographic Parity. In addition, the CDP aligns with Fairness through Unawareness when all variables are allowed in the conditioning set. This is a concept defined later in Section~\ref{FTU}.

\subsubsection{Equality of Opportunity (EOO)}
A predictor $\hat{Y}$ meets Equality of Opportunity if the expected predicted pure premiums are equal across groups after controlling for the actual risk $Y$. This is expressed as: $$E(\hat{Y}|Y=y,D=a)=E(\hat{Y}|Y=y,D=b).$$
EOO for continuous outcomes requires the expected predicted risk score to be equal across groups defined by a protected attribute, conditional on the true underlying loss. This means that among individuals with the same expected claim cost, the model’s predicted risk must not systematically vary by sensitive characteristics \citep{roemer2015equality}. When the true risk cost $Y$ is fully determined by non-protected variables $X$, EOO is equivalent to Conditional Demographic Parity (CDP).

\subsection{Individual Fairness}
\subsubsection{Fairness through Unawareness (FTU)}\label{FTU}
A predictor $\hat{Y}$ achieves Fairness through Unawareness if the premium calculation excludes the sensitive variable $D$ $$\hat{Y}=\mu(X).$$
The application of FTU eliminates direct discrimination by prohibiting the use of sensitive attributes such as gender and race in the premium-setting process. Individuals with identical non-sensitive characteristics receive the same premium regardless of their group membership under this principle, which provides an intuitive sense of fairness. In practice, FTU is the most straightforward fairness criterion to implement and is often realized by simply not collecting sensitive information at all. This is a common approach in regulatory frameworks. For instance, the EU Gender Directive bans the use of gender in insurance pricing, which causes many EU insurers to refrain from collecting gender data from their policyholders. However, this approach can have unintended consequences: the use of unisex actuarial tables in life or health insurance may obscure genuine risk differences. This could potentially introduce bias, distort risk pools, and encourage adverse selection. These issues are particularly problematic when genders are not balanced among policyholders \citep{chen2017unisex}.

\subsubsection{Fairness through Awareness (FTA)}\label{fta}
Fairness through awareness is met if similar individuals get similar premium predictions. This is expressed as: $$\hat{Y}(X^{(i)},D^{(i)})\approx\hat{Y}(X^{(j)},D^{(j)}) \ \ \ \text{if}\ \ \ \ d(X^{(i)},X^{(j)})\ \ \text{ is small}.$$
FTA is an individual-level fairness concept introduced by \cite{dwork2012fairness}, which requires that similar individuals receive similar treatment. This criterion is used by enforcing a Lipschitz constraint during the loss minimization process, which ensures that the difference in predicted pure premiums between two individuals is bounded by their distance in a predefined similarity metric: $$\forall x, y \in \mathcal{X} \times \mathcal{X}: \Delta(\mu(x),\mu(y))\le L\cdot d(x,y)+\tau,$$ where $d$ and $\Delta$ denote two specific measures of similarity of distributions and $\tau>0$ is a small constant. FTA is often argued to be incompatible with DP.

One challenge in implementing FTA lies in the specification of the dissimilarity metric $d:\mathcal X\times\mathcal X \mapsto \mathbb{R}_+$. 
While \cite{dwork2012fairness} argued that the metric should be designed using domain expertise and human judgment to reflect task-specific notions of similarity, 
defining such a metric in practice can be subjective and difficult to implement.

   

In the context of insurance pricing, the actuarial risk score that represents an individual’s expected loss can serve as a domain-appropriate measure of similarity between policyholders \citep{dwork2012fairness}. Thus, this score may function as a natural metric $d(x,y)$ to assess proximity or comparability between individuals, which facilitates the implementation of fairness through awareness (FTA). Regulatory bans on sensitive attributes encourage Fairness Through Awareness (FTA) by directing insurers to base pricing on predictive and non-protected variables that better approximate true risk. This promotes equitable treatment while preserving model performance. 

The Lipschitz constant serves as a mechanism for individual fairness by bounding the sensitivity of predictions to small perturbations in features, which directly formalizes FTA. However, estimating a global Lipschitz constant is often computationally heavy and susceptible to bias in sparse or unrepresentative regions of the data \citep{petersen2021post}. We thus adopt a local Lipschitz constant, which is defined as the 95th percentile of local sensitivity ratios:
\[
\text{Local Lipschitz Constant} \;=\; Q_{0.95}\!\left( \frac{\big|mu(x_i) - \mu(x_{i\text{-}NN})\big|}{d(x_i, x_{i\text{-}NN})} \right),
\]
where $x_{i\text{-}NN}$ is the nearest neighbor of $x_i$ in non-protected features. $Q_{0.95}$ denotes the 95th percentile. $d(\cdot,\cdot)$ is the Gower distance, which is a robust metric for mixed-type risk profiles \citep{gower1971general}. This yields a scalable, context-aware fairness assessment at the individual level.

\subsubsection{Counterfactual Fairness (CF)}\label{cf}
A predictor $\hat{Y}$ is counterfactually fair if 
$$\hat{Y}_{D\gets a}(X,D=a)=\hat{Y}_{D\gets b}(X,D=a).$$
The concept of counterfactual fairness originates from Pearl's causal model \citep{pearl2009causality} and answers a specific question: if a female were male in a counterfactual world, would the premium they receive change? 
This notion was introduced by \cite{kusner2017counterfactual} and states that the sensitive attribute should not causally affect the premium for any individual. It is also important to note that CF and FTA are related but distinct. Although FTA places more emphasis on individual similarity and offers a general framework for individual-level fairness, CF strengthens it by incorporating causality. This ensures that similarity is not just statistical but ethically justified. 

There have been significant methodological advancements in approaches to counterfactual fairness (CF) in machine learning \citep{kusner2017counterfactual,kilbertus2017avoiding}. 
However, counterfactual fairness remains underexplored in the insurance industry. This is due not only to the challenges of specifying a well-defined causal model that accurately captures the complex relationships among variables but also to the difficulties in quantifying counterfactual outcomes and translating them into enforceable regulatory standards. Although causal graphs and underlying variable relationships are increasingly acknowledged in the development of fair pricing models in insurance, the primary focus remains on individual and group fairness rather than on counterfactual fairness \citep{cote2025fair}.

We propose here a novel metric to quantify counterfactual fairness. 
Counterfactual Fairness is naturally evaluated using the individual treatment effect (ITE) since it operates at the individual level. A practical and efficient approach to estimating ITE is causal forests \citep{athey2019estimating}, which use random forest splitting to adapt to heterogeneity and estimate treatment effects within leaf nodes. This method performs consistently well in high-dimensional and large-scale observational settings. This makes it well-suited for fairness auditing in insurance \citep{wager2018estimation}. We enforce \emph{honest} splitting and fix the random seed to ensure reproducibility.

As ITE distributions in real-world data are often asymmetric, we adopt the \emph{median ITE} as our main summary fairness metric for its robustness. We also examine the full ITE distribution to assess fairness heterogeneity across populations. Let $L$ denote a leaf (terminal) node in the forest. The ITE for $L$ is the difference in group-averaged outcomes:
\[
\text{ITE}_L
= \frac{1}{|L_a|}\sum_{i \in L_a} \hat Y_i \;-\; \frac{1}{|L_b|}\sum_{j \in L_b} \hat Y_j,
\] 
where $L_a = \{i: D_i = a, x_i \in L\}$ and $L_b = \{j: D_j = b, x_j \in L\}$.
The overall counterfactual fairness metric is then $Q_{0.5}(\text{ITE}_L)$, i.e., the median across all leaves.

\section{Methodology}\label{sec:method}
\subsection{Predictive Accuracy}
In order to evaluate competing pricing models, we consider both fairness and predictive accuracy as key criteria. Fairness across multiple dimensions including group, individual, and counterfactual fairness has been rigorously quantified in the preceding Section~\ref{def}. However, predictive accuracy remains essential for actuarial soundness and market competitiveness.

In terms of predictive accuracy, we evaluate model performance using the Root Mean Square Error (RMSE) and the Normalized Gini Index. These metrics are defined as follows:
{\allowdisplaybreaks
\begin{align*}
\text{RMSE} &= \sqrt{\frac{1}{n} \sum_{i=1}^{n} \left( y_i - \hat{y}_i \right)^2}, \\
\text{Normalized Gini Index} &= \frac{\frac{\sum_{i=1}^ny_ir(\hat{y}_i)}{\sum_{i=1}^ny_i}-{\sum_{i=1}^n\frac{n-i+1}{n}}}{\frac{\sum_{i=1}^ny_ir(y_i)}{\sum_{i=1}^ny_i}-\sum_{i=1}^n\frac{n-i+1}{n}},
\end{align*}
where $ y_i $ is the true outcome; $ \hat{y}_i $ is the predicted value for the $i$-th observation; $ n $ is the sample size; $r$ denotes the rank of an individual’s actual claim amount $ y_i $ or predicted claim $ \hat{y}_i $ within the population and is ordered from lowest to highest. The Normalized Gini Index measures the rank correlation between predictions and actual outcomes. This emphasizes the model’s ability to correctly order risks, which is a key consideration in insurance pricing \citep{ye2022combining}.
}
\subsection{Models}\label{model}
In this section, we discuss various cost model designs that aim to eliminate discrimination and satisfying the fairness criteria defined in Section~\ref{def}. It is important to clarify that the ``cost model'' here refers specifically to the pure premium prediction model, which estimates expected claim costs. For the broader pricing process including price optimization and demand modeling, we refer readers to \citet{shimao2025welfare}. Based on how fairness is enforced, we categorize models into three paradigms: \textit{preprocessing} (modifying the input data prior to training), \textit{inprocessing} (incorporating fairness constraints during model training), and \textit{postprocessing} (adjusting predictions after model output) \citep{kamiran2013techniques}. In the empirical analysis presented in Section~\ref{analysis}, each fairness approach is implemented within two distinct predictive frameworks. First, we use a traditional Generalized Linear Model (GLM) with Poisson and Gamma distributions for claim frequency and severity, respectively. Second, we use a machine learning model using XGBoost \citep{chen2016xgboost}. We select XGBoost due to its superior performance in handling large datasets and its enhanced predictive accuracy compared to other ensemble learning methods. Both are used to capture the relationship $\mu$ between policyholder characteristics and claim outcomes. All notation is consistent with those stated in Section \ref{def}.\\

\textbf{BEST ESTIMATE PRICE (MB)} refers to the prediction generated by a full model that incorporates all available covariates, including both protected and non-protected variables. This prediction represents the most accurate estimate of the expected claim cost since it uses the complete set of predictive information. However, such a model raises concerns regarding both \textit{direct discrimination} and \textit{indirect discrimination} while maximizing predictive performance. The model MB is expressed as:
$$\hat{Y}=\mu(X,D).$$

\textbf{UNAWARENESS MODEL (MU)} removes the sensitive attribute directly from the feature set in the preprocessing stage, which implements the principle of FTU discussed in Section~\ref{FTU}. While this approach eliminates explicit use of sensitive variables, it fails to address indirect discrimination. This is due to the fact that the model may still exploit proxy variables to achieve differential treatment. Proxy variables are features highly correlated with the sensitive attribute. Consequently, disparities can persist even in the absence of direct attribute inclusion. The model is expressed as:
$$\hat{Y}=\mu(X).$$

\textbf{ORTHOGONAL MODEL (MO)} is a preprocessing approach that modifies the training data prior to model fitting. It is intended to reduce dependence on $D$, and thus improve DP. It eliminates indirect discrimination by removing components of non-protected features that are correlated with the sensitive attribute. This is achieved through residualization: each non-protected variable is regressed on the sensitive attribute and replaced by its residual. This technique effectively isolates the part of the feature that is linearly uncorrelated with group membership. 

In our implementation, we assume a linear relationship between the sensitive attribute and the non-protected variables. We use ordinary least squares regression to perform the adjustment. 
The model is expressed as:
$$\hat{Y}=\mu(X^*),$$ 
$$X^*=X-\hat{X},$$
$$\hat{X}=\textbf{1}\cdot b_0+D\cdot b_1.$$

\textbf{DISCRIMINATION-FREE MODEL (MDF)} is a model that is consistent with DF and originally proposed by \citet{pope2011implementing}. It is a postprocessing method that constructs fair predictions by averaging the best-estimate prices over a specified distribution of the sensitive attribute. As suggested by \citet{lindholm2022discrimination}, a natural and practical choice is the empirical distribution of the sensitive variable. In the case of a binary sensitive attribute (e.g., gender), this corresponds to weighting the counterfactual predictions by the observed group proportions in the population. This approach ensures that the resulting premium is independent of the individual's sensitive attribute while preserving predictive accuracy on average, and it is probabilistically justified as a form of fair averaging. The MDF model is expressed as:
$$\hat{Y}=\int \mu(X,D)dP(D)=\mu(X,D=a)\cdot \mathbb{P}(D=a)+\mu(X,D=b)\cdot \mathbb{P}(D=b).$$

\textbf{BARYCENTER MODEL (MBC)} is a postprocessing method introduced by \citet{charpentier2024insurance} and designed to achieve group fairness while preserving individual risk sensitivity. It uses the concept of \emph{barycenters} from optimal transport theory to adjust predictions in a way that balances fairness and accuracy \citep{chzhen2020fair}. The method aligns the predictive distributions across groups by transforming the risk scores of one group to match the distributional characteristics of the other. It ensures that both groups have the same marginal distribution of adjusted predictions, thereby satisfying demographic parity in expectation.
For individuals in group $D=a$, the model retains their direct prediction $\mu(X,D=a)$ weighted by $\mathbb{P}(D=a)$. In addition, the model adds a counterfactual component derived by mapping their risk score through the inverse cumulative distribution function of the other group. This is expressed mathematically as $F^{-1}_B \circ F_A(\mu(X,D=a))$, where $F_A$ and $F_B$ are the empirical CDFs of predicted risks for groups $a$ and $b$ respectively. This transformation ensures that no group systematically receives higher or lower premiums, while maintaining rank consistency within groups. MBC is expressed as:
\begin{align*}\hat{Y}(X,D=a)&=\mathbb{P}(D=a)\cdot \mu(X,D=a)+\mathbb{P}(D=b)\cdot F^{-1}_B\circ F_A(\mu(X,D=a)).\\ \hat{Y}(X,D=b)&=\mathbb{P}(D=b)\cdot \mu(X,D=b)+\mathbb{P}(D=a)\cdot F^{-1}_A\circ F_B(\mu(X,D=b)).\end{align*}

\textbf{SYNTHETIC CONTROL METHOD MODEL (MSCM)} is a preprocessing method that adjusts the claim amount rather than directly modifying customer characteristics. We construct a debiased target by taking a weighted average of the actual claim and a counterfactual estimate to mitigate potential gender-related bias in observed claims. The counterfactual estimate represents the claim that would have occurred under a different gender assignment, holding other risk factors constant. Inspired by the synthetic control framework of \citet{abadie2010synthetic}, we employ SCM to generate these counterfactual claims. This approach offers a data-driven, transparent, and interpretable method for estimating the scenario that occurs in the absence of a specific treatment or attribute. In addition, this approach implicitly accounts for unobserved confounders through pretreatment outcome matching. By training models on these adjusted claims, we aim to reduce the causal influence of gender on pricing without sacrificing predictive accuracy too much. 

A critical limitation of this approach arises from the nature of the sensitive attribute. As gender is time-invariant, there is no pretreatment period in which to evaluate the fit of the synthetic control. For standard synthetic control applications, an intervention occurs at a defined point in time and pre-intervention outcomes enable model validation. However, the attribute-based "treatment" considered here is inherent and permanent. This precludes conventional placebo tests based on pretreatment parallelism or predictive accuracy. Consequently, the validity of the estimated counterfactual claims depends on the assumption that the model correctly captures the underlying risk structure across gender groups. For readers interested in alternative approaches, methodological developments such as the robust synthetic control of \citet{amjad2018robust} and the augmented synthetic control method of \citet{ben2021augmented} provide frameworks that relax temporal assumptions and improve estimation robustness in non-traditional causal settings. MSCM is formulated as:
\begin{align*}
    \hat{Y}&=\mu(X,Y'),\\
    Y'&=\frac{Y_0+Y_{counterfactual}}{2},\\
    Y_{counterfactual}&=Y_1^\top\cdot W,\\
    W&=\argmin_W  ||X_0-X_1W||=\argmin_W  \sqrt{(X_0-X_1W)'V(X_0-X_1W)},
\end{align*}
where $Y_0$ and $X_0$ denote the claim outcome and non-protected characteristics of a specific individual respectively; $Y_1$ and $X_1$ represent the corresponding claim outcomes and non-protected features of the risk pool consisting of policyholders with the opposite value of the sensitive attribute. The weight matrix $V$ reflects the relative importance of non-protected variables in predicting risk. In our implementation, the weights are derived from the feature importance scores of a trained random forest model, which provides a non-linear and data-driven assessment of the predictive contribution of each variable. 

\textbf{TWO-BRANCH NEURAL NETWORK MODEL (MNN)} is specifically designed to obtain suitable counterfactual fairness, unlike the previously discussed models, which are applied to both XGBoost and GLM estimation models. 
The architecture consists of a shared lower network that learns common feature representations, followed by two separate output heads: one for the real-world prediction and one for the counterfactual-world prediction. For each individual, the model processes both their observed features and their counterfactual counterpart (e.g., with the sensitive attribute flipped from MSCM) through the shared layers. This allows the network to learn invariant representations that are robust to changes in the sensitive attribute.

The model is trained using a composite loss function that jointly optimizes for predictive accuracy and counterfactual fairness:
\[
\mathbf{L} = \frac{1}{n}\sum_{i=1}^n \left(Y_i - f^{\text{real}}(x_i)\right)^2 + \lambda \cdot \frac{1}{n}\sum_{i=1}^n \left(f^{\text{real}}(x_i) - f^{\text{counterfactual}}(x_i')\right)^2,
\]
where $f^{\text{real}}(x_i)$ is the prediction for the individual in the real world, and $f^{\text{counterfactual}}(x_i')$ is the prediction for their counterfactual version; $\lambda > 0$ is a hyperparameter that controls the trade-off between accuracy and fairness. A higher $\lambda$ enforces greater invariance to the sensitive attribute, and thus enhances counterfactual fairness. The optimal value of $\lambda$ is determined via 5-fold cross-validation, as detailed in Appendix~\ref{lambda}.

\section{Empirical Analysis} \label{analysis}
\subsection{Datasets}
In this section, we conduct an empirical analysis using two private motor insurance datasets from the \texttt{CASDataset} in R \citep{dataset}. A detailed description of all variables in the two datasets is provided in Appendix~\ref{dataset}.

The first dataset, \textit{pg15training}, contains 100,000 third-party liability (TPL) policies. It includes claims data on both material damage and bodily injury. We focus on material claims due to their higher claim frequency. Gender is treated as the sensitive attribute, while non-protected variables are provided in Table~\ref{tpg} in Appendix \ref{dataset}. 

The second dataset, \textit{fremotor1prem0304a}, consists of 17,001 complete policy records with claim amounts from 2003 and 2004. The standard frequency-severity modeling approach cannot be applied since claim count information is not available in this dataset. Instead, we model the total claim amount directly using a Gamma Generalized Linear Model (GLM), which is well-suited for positive, continuous, and right-skewed responses such as insurance claims \citep{goldburd2016generalized}. The total claim amount covering all guarantees is used as the outcome variable. The sensitive attribute is again gender, and non-protected covariates are provided in Table~\ref{tfre} in Appendix \ref{dataset}. Additionally, we construct an insurance risk score using selected risk factors to capture unobserved risk heterogeneity. These factors are also shown in Table~\ref{tfre} in Appendix \ref{dataset}.

\subsection{Model Comparison}\label{comparison}
We evaluate all proposed models on both datasets using the metrics defined before. To facilitate comparison, we employ trade-off visualizations that plot predictive accuracy against a selected fairness criterion. Each point represents a model's performance, which allows for a transparent assessment of the balance between accuracy and fairness. This approach enables practitioners to select models based on their desired trade-offs, particularly in regulated or ethically sensitive contexts such as insurance pricing.

\subsubsection{Group Fairness and Accuracy}
\begin{figure}[h!]
    \centering
    \includegraphics[width=0.95\linewidth]{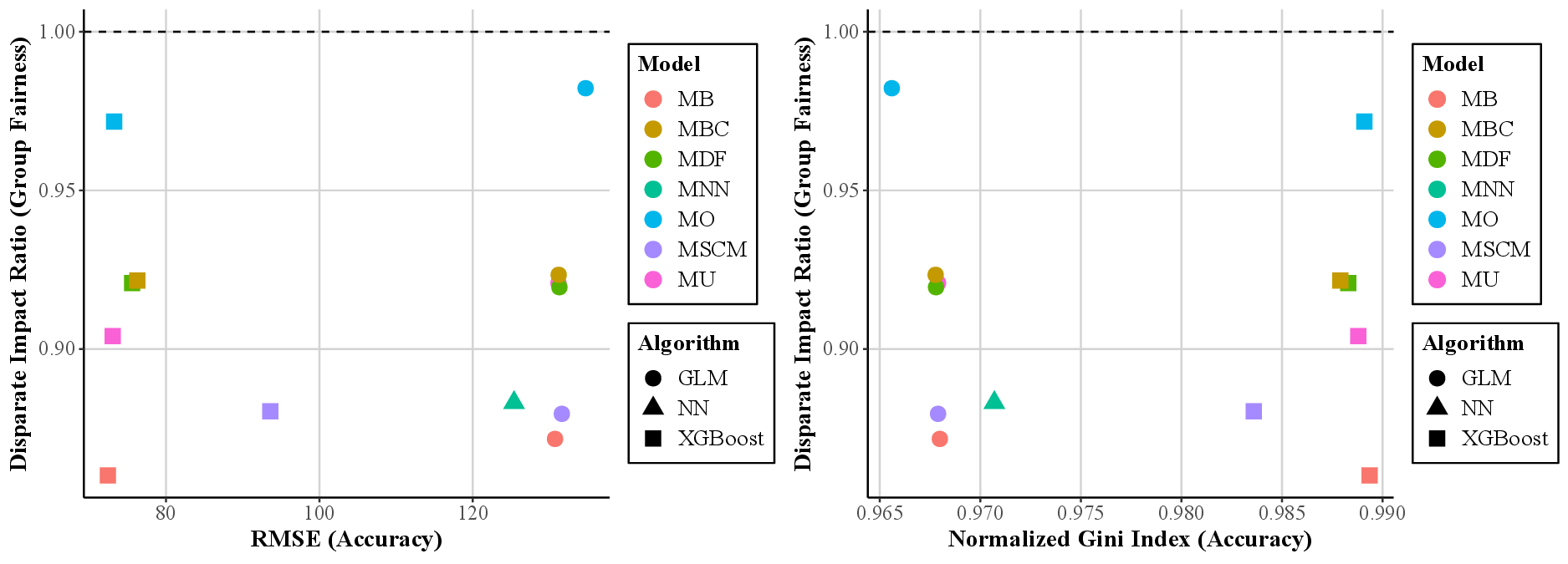}
    \caption{Group fairness-accuracy plot  (\textit{fremotor1prem0304a})}
    \label{FRE GF}
\end{figure}

\begin{figure}[h!]
    \centering
    \includegraphics[width=0.95\linewidth]{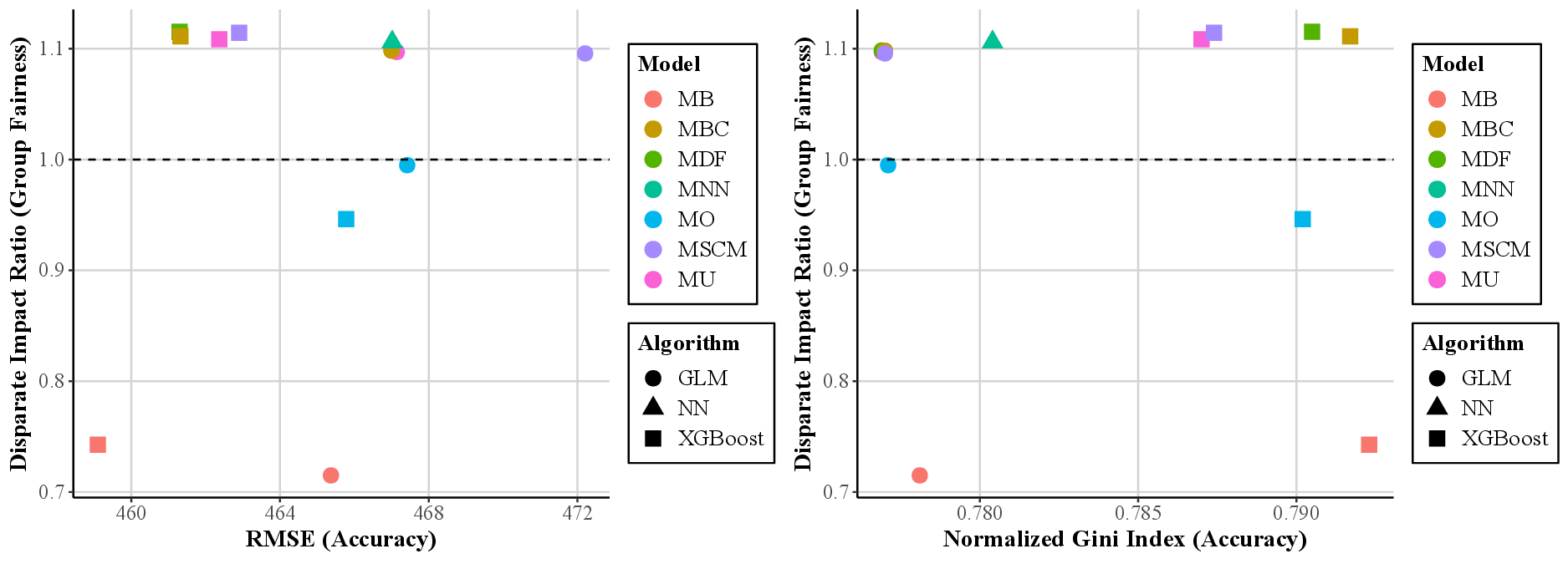}
    \caption{Group fairness-accuracy plot (\textit{pg15training})}
    \label{PG GF}
\end{figure}

Figures~\ref{FRE GF} and~\ref{PG GF} illustrate the trade-off between prediction accuracy and demographic parity across the evaluated models. In both datasets, XGBoost generally yields more accurate predictions than GLM. The MB model, which includes all available features, achieves the highest predictive accuracy (lowest RMSE), but exhibits the worst group fairness. This is due to the model consistently predicting higher average premiums for male policyholders. When gender is excluded from the MU model, a significant shift in predicted premiums is observed: females face higher predicted costs while males experience reductions. This indicates that gender carries predictive power for risk. The removal of it disrupts the model’s ability to capture true risk differences, which leads to substantial reallocations. 


It is worth noting that the adjustment for the \textit{pg15training} dataset is so pronounced that the gender-based premium ratio reverses—shifting from below 1 (favoring males) to above 1 (favoring females). This over-correction suggests that the model compensates for the missing gender variable by relying on correlated proxies (age in our case), which results in an excessive and potentially unfair redistribution of risk costs. This highlights the limitations of simple unawareness as a fairness strategy in the presence of strong proxy variables. The MO model achieves the best performance in terms of group fairness especially with GLM, which results in the most similar average premiums across gender groups. This suggests that the underlying risk distribution is approximately equal across genders after adjusting for observable risk factors. Besides this, the observed disparities under the baseline model are primarily driven by discriminatory pathways rather than genuine risk differences. The resulting alignment of group averages after orthogonalization supports the effectiveness of MO in promoting equitable outcomes.

In contrast, MDF and MBC models exhibit similar performance since both operate through postprocessing reweighting of predictions. While these methods improve fairness compared to the baseline, their ability to correct group-level disparities is limited. This is because they do not modify the model’s internal dependence on proxy variables. MSCM performs worse than the MU model in group fairness, and it incurs a noticeable loss in predictive accuracy. This degradation suggests that the counterfactual claim adjustment used in MSCM may introduce noise or bias, which makes it less suitable for achieving a favorable trade-off between group fairness and performance in this context.

\subsubsection{Individual Fairness and Accuracy}\label{if}
\begin{figure}[htbp]
    \centering
    \includegraphics[width=0.95\linewidth]{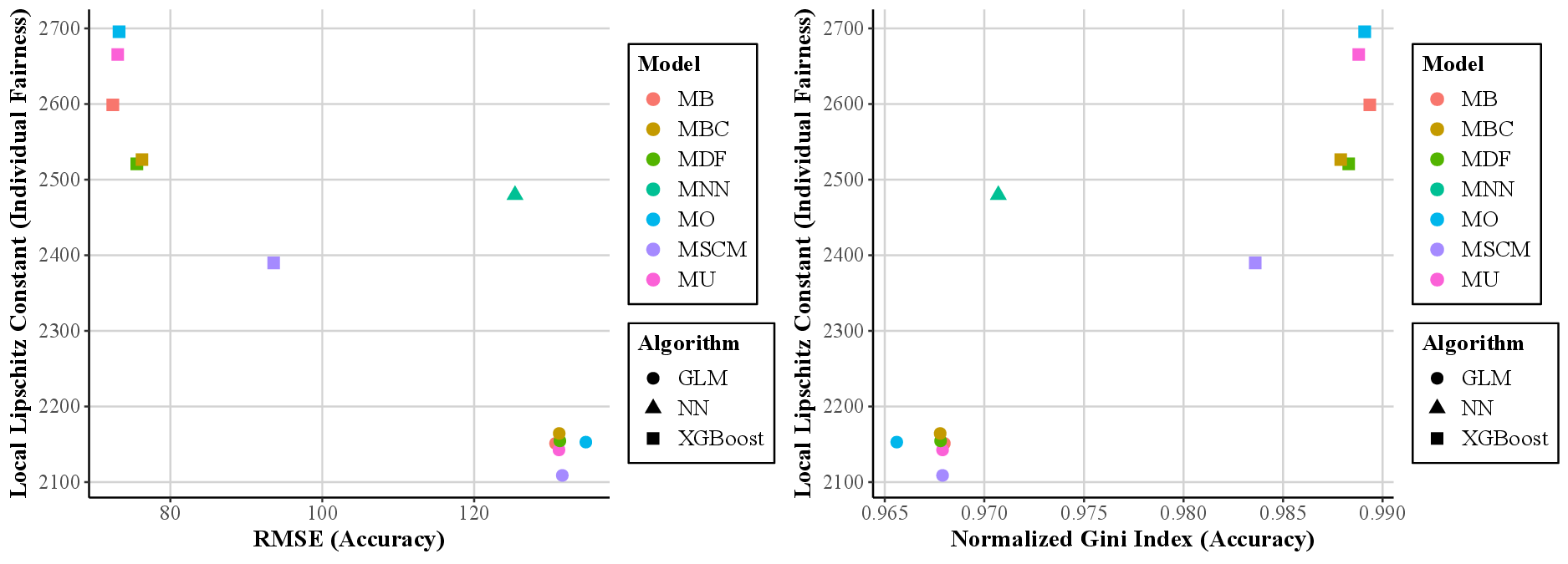}
    \caption{Individual fairness-accuracy plot (\textit{fremotor1prem0304a})}
    \label{FRE IF}
\end{figure}

\begin{figure}[htbp]
    \centering
    \includegraphics[width=0.95\linewidth]{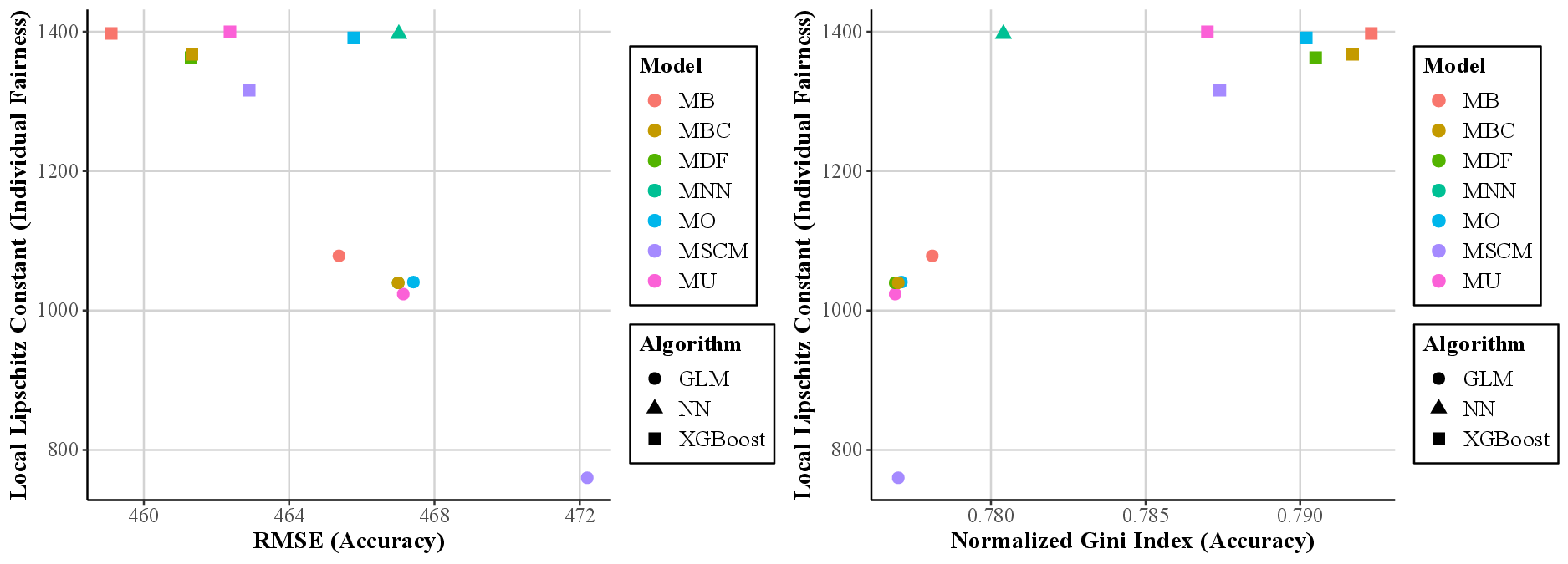}
    \caption{Individual fairness-accuracy plot (\textit{pg15training})}
    \label{PG IF}
\end{figure}

MU, MSCM, and MDF are expected to satisfy individual fairness by design, as they aim to decouple predictions from sensitive attributes. However, their empirical performance can vary significantly in practice, depending on the choice of the similarity metric \(d(x, x')\) and the underlying data distribution \citep{agarwal2021towards}. As shown in Figures~\ref{FRE IF} and~\ref{PG IF}, GLM-based models generally exhibit better individual fairness compared to their XGBoost counterparts. This is attributed to the inherent linearity and smoothness of GLMs, which produce more stable and continuous predictions. In contrast, tree-based models such as XGBoost are piecewise constant and more sensitive to small perturbations in input features. This leads to larger local Lipschitz constants and greater potential for unfair treatment of similar individuals \citep{ranzato2021fairness}. 

MU under XGBoost demonstrates a particularly high Local Lipschitz constant, i.e., large changes in predicted premium for small changes in input features, and this indicates a pronounced violation of individual fairness. Although MU excludes the sensitive attribute, this occurs due to the tree-based structure of XGBoost readily exploiting complex and non-linear interactions among other features that serve as proxies for the sensitive attribute. While MB uses the sensitive attribute directly and applies it in a relatively consistent and smooth manner, MU forces the model to approximate that information indirectly. This results in unstable and non-smooth decision boundaries. As a consequence, even small differences in input features can trigger large and discontinuous changes in predicted premiums. This is particularly observed in regions where the splitting behavior is most influenced by these proxy correlations \citep{vargo2021individually}. This inherent instability of tree ensembles when handling masked sensitive information underscores a key limitation of simple unawareness in non-linear and high-capacity models.

MBC and MDF show comparable performance since both preserve local prediction smoothness. However, MO exhibits poorer individual fairness. This degradation arises because MO modifies the non-protected features through residualization, which distorts the original feature space. As a result, the identification of the nearest neighbors becomes less meaningful, which erodes the validity of the distance metric and compromises the model’s ability to treat similar individuals similarly. 

\hspace{0.0em}

\subsubsection{Counterfactual Fairness and Accuracy} 
By design, MSCM and MNN significantly improve counterfactual fairness by pushing the metric towards 0. MNN proves to be an effective approach with minimal compromise on predictive accuracy when achieving counterfactual fairness is the only priority. In fact, MNN’s predictive performance falls between that of the highly flexible XGBoost and the more rigid GLM. This strikes a favorable balance between fairness and model utility. It is natural for MB to exhibit a large treatment effect due to direct use of gender, but the poor counterfactual fairness of MO is counterintuitive. The issue arises because MO modifies features via residualization, but the tree splitting is still based on the original unadjusted variables. This creates a mismatch: the original features capture a more comprehensive risk profile, while the debiased features used for prediction eliminate some risk associations. As a result, individuals in the same leaf node (based on raw features) may have divergent debiased representations. This leads to larger differences in predicted outcomes, which inflates the estimated treatment effect and compromises MO’s counterfactual fairness goal.

MDF performs similarly to MU, and both models outperform MBC. Although MBC and MDF achieve comparable levels of group and individual fairness, they diverge in counterfactual fairness due to MBC’s use of optimal transport to align prediction distributions. This technique balances marginal prediction distributions across groups, but does so by re-scaling individual predictions based on within-group rank instead of their risk profiles. This creates a critical issue: two individuals with similar characteristics may receive very different adjustments depending on their group’s overall performance distribution. The causal forest thus detects artificial disparities in the counterfactual space, which inflates the estimated treatment effect for MBC.
\begin{figure}[H]
    \centering
    \includegraphics[width=0.95\linewidth]{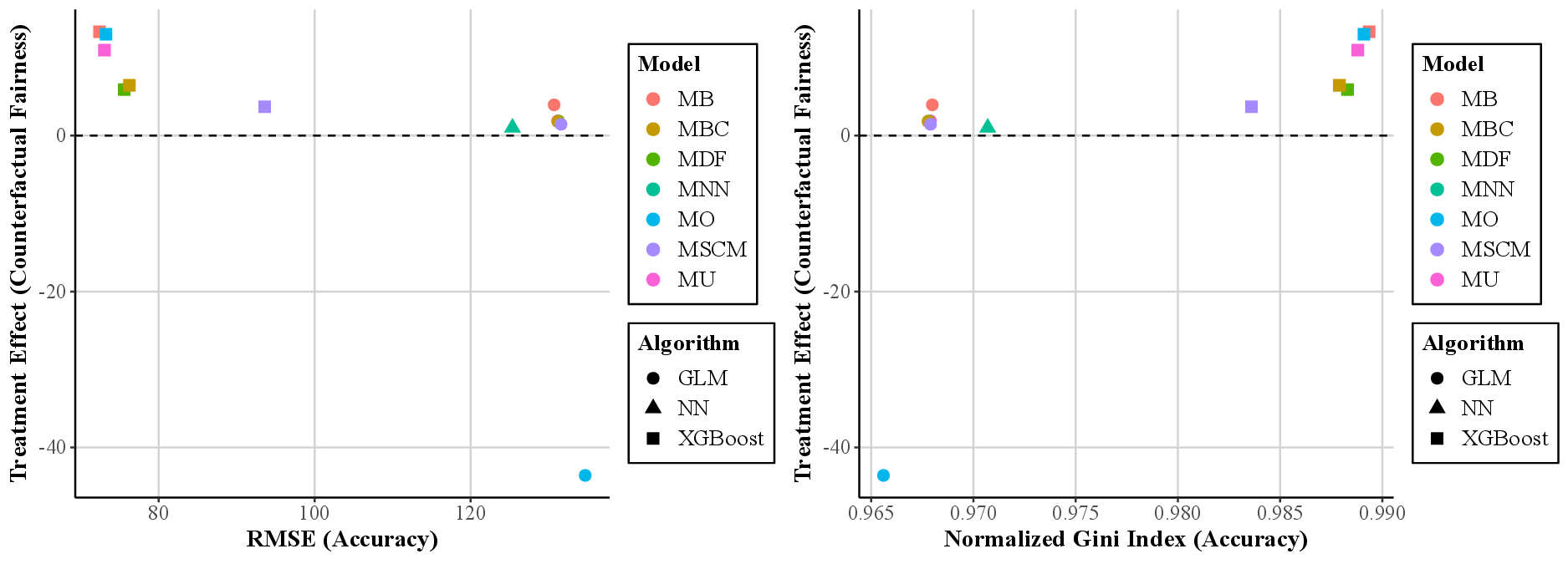}
    \caption{Counterfactual fairness-accuracy plot (\textit{fremotor1prem0304a})}
    \label{FRE CF}
\end{figure}
\begin{figure}[H]
    \centering
    \includegraphics[width=0.95\linewidth]{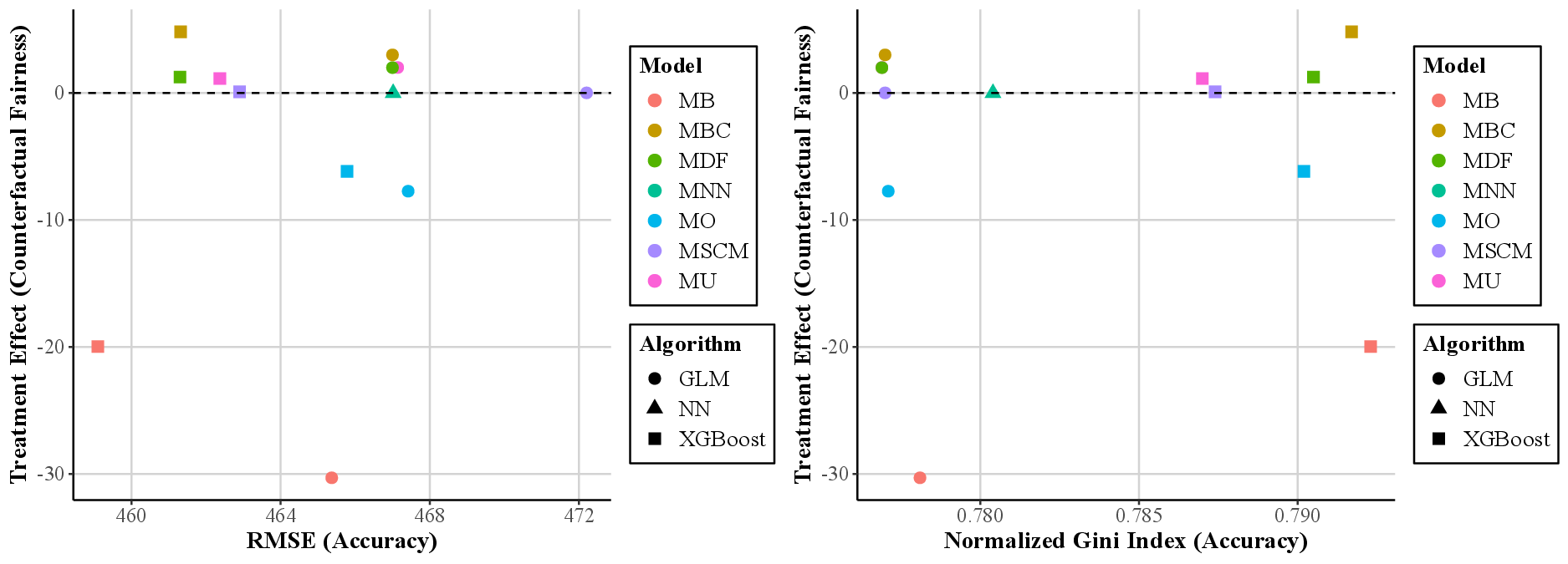}
    \caption{Counterfactual fairness-accuracy plot (\textit{pg15training})}
    \label{PG CF}
\end{figure}

\subsection{Counterfactual Performance}
As discussed in Section~\ref{cf}, summarizing counterfactual performance using the median of the Individual Treatment Effect (ITE) may be misleading. This is because it fails to capture distributional characteristics such as skewness, multimodality, or heavy tails. To gain a more comprehensive understanding, we analyze the full density distribution of ITE across models. A distribution with a sharp peak centered near zero and narrow tails indicates that the treatment effect is consistently close to zero across individuals—reflecting strong counterfactual fairness. Conversely, broader or bimodal distributions suggest heterogeneous effects and weaker fairness guarantees. This visualization allows us to assess not only the central tendency but also the stability and uniformity of fairness across the population.
\begin{figure}[H]
    \centering
    \includegraphics[width=0.95\linewidth]{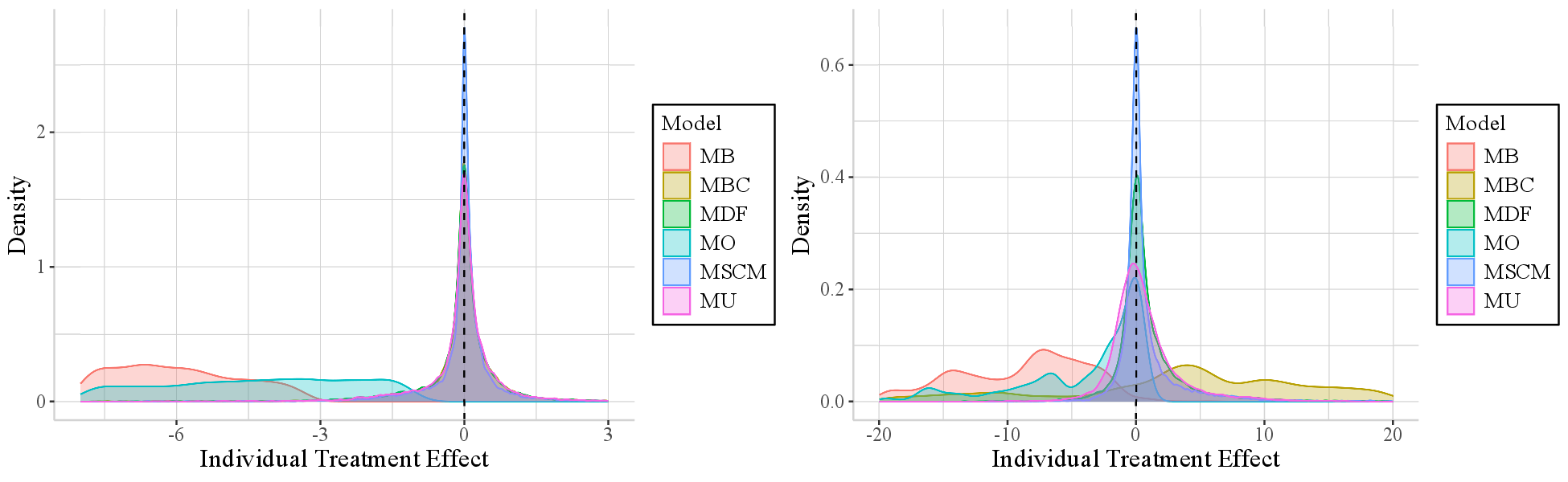}
    \caption{Density Plot of ITE in \textit{pg15training} (Left: GLM, Right: XGBoost)}
    \label{pgcf}
\end{figure}
\begin{figure}[H]
    \centering
    \includegraphics[width=0.95\linewidth]{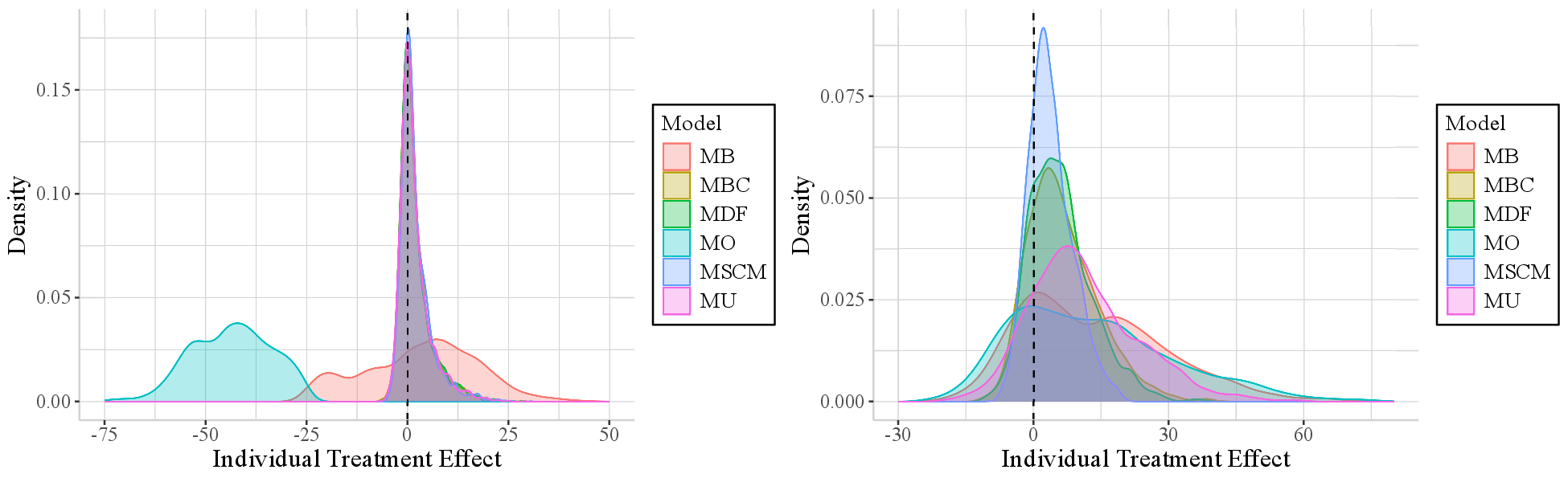}
    \caption{Density Plot of ITE in \textit{fremotor1prem0304a} (Left: GLM, Right: XGBoost)}
    \label{frecf}
\end{figure}
As shown by Figures~\ref{pgcf} and \ref{frecf}, MSCM consistently produces the sharpest density peak centered around zero with significantly narrower tails compared to other models. This indicates superior performance in counterfactual fairness as the distribution of ITE is tightly concentrated near zero across the population. In this case, the median of the ITE serves as a reliable and informative summary of the overall distribution. This reflects the model’s consistent mitigation of gender-based disparities. To further evaluate the effectiveness of different counterfactual modeling strategies, we compare MSCM with MNN. We plot the density of ITE for MNN, MSCM with GLM, and MSCM with XGBoost.

\begin{figure}[htbp]
\centering 
\subfigure{
\label{Fig.sub.1}
\includegraphics[width=0.485\linewidth]{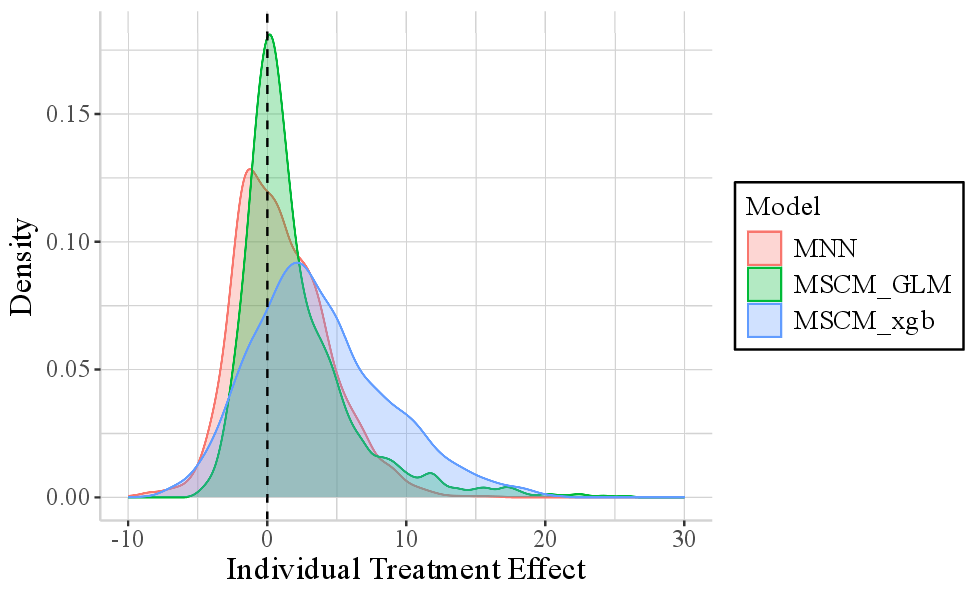}}
\subfigure{
\label{fre}
\includegraphics[width=0.485\linewidth]{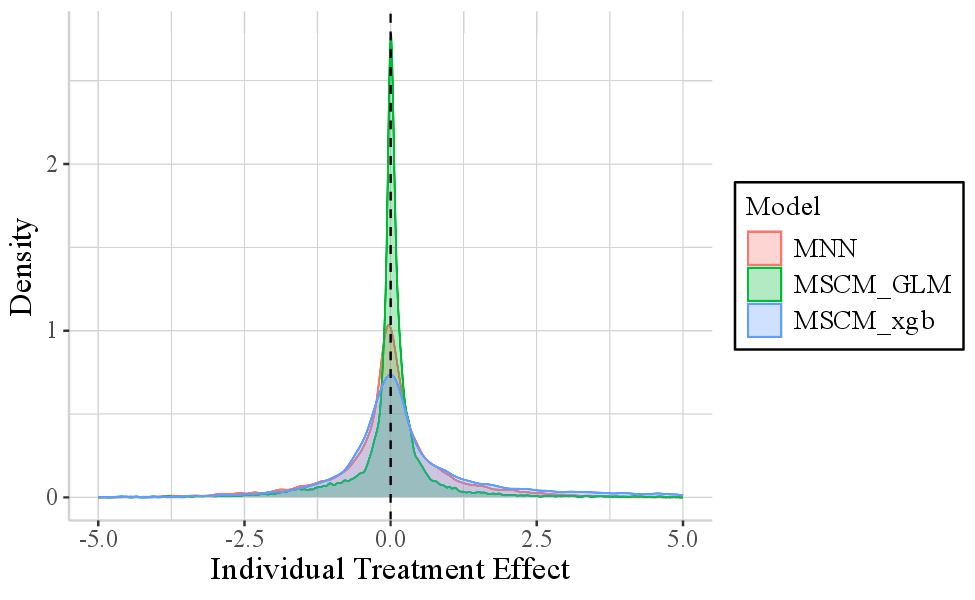}}
\caption{Density plot of ITE (Left:\textit{fremotor1prem0304a}, Right: \textit{pg15training})}
\label{pg}
\end{figure}
Based on Figure~\ref{pg}, MNN achieves a counterfactual fairness metric closer to zero than MSCM. However, its overall performance across the full distribution of ITE is not optimal. In both datasets, the distributional performance of  MNN lies between that of MSCM-GLM and MSCM-XGBoost. The MSCM-XGBoost exhibits the sharpest peak around zero and the narrowest tails, which indicates superior counterfactual fairness. Given that MNN relies on a more complex and instance-based matching mechanism that reduces model interpretability and increases computational cost, its marginal gain in central tendency does not justify its added complexity. In contrast, MSCM achieves excellent counterfactual fairness through a principled and transparent adjustment based on synthetic control. This makes it more robust and interpretable. Therefore, MSCM emerges as the preferred method over MNN for achieving counterfactual fairness in insurance pricing when both distributional performance and practicality are considered. 

\subsection{Solidarity and Adverse Selection}\label{sec:sol}
\subsubsection{Solidarity}
When discussing group fairness in insurance, a closely related concept is \textit{solidarity}. This refers to the collective sharing of risk and financial responsibility within an insurance pool. This principle often manifests through cross-subsidization, which is a scenario where lower-risk individuals effectively subsidize premiums for higher-risk individuals. This promotes inclusivity and equitable access to coverage.

In our analysis, we investigate the presence and extent of cross-subsidization with respect to two key demographic factors: age and gender. We follow the framework of \citet{henckaerts2021boosting} and assess how fairness-motivated models redistribute risk compared to a benchmark model using the difference $Y_{fair}-Y_{benchmark}$. We select the Best Estimate model (MB) that maximizes predictive accuracy by using all available information as our benchmark. Deviations from MB's predictions are interpreted as the price of fairness, which reflects the degree of redistribution induced by each fair model.

\begin{figure}[H]
    \centering
    \includegraphics[width=0.70\linewidth]{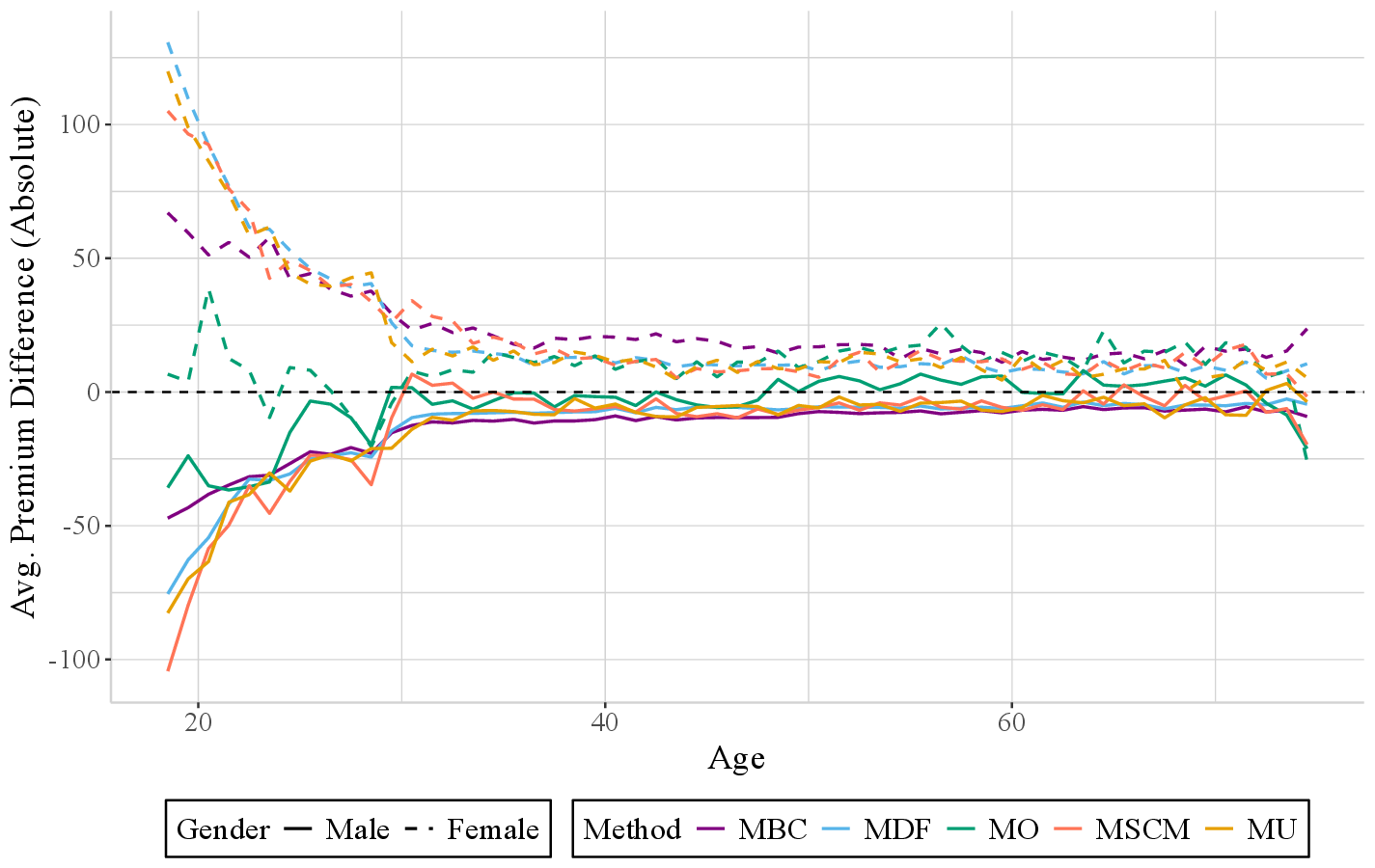}
    \caption{Absolute premium difference in \textit{pg15training} (XGBoost models versus XGBoost MB)}
    \label{pgsol}
\end{figure}
\begin{figure}[H]
    \centering
    \includegraphics[width=0.70\linewidth]{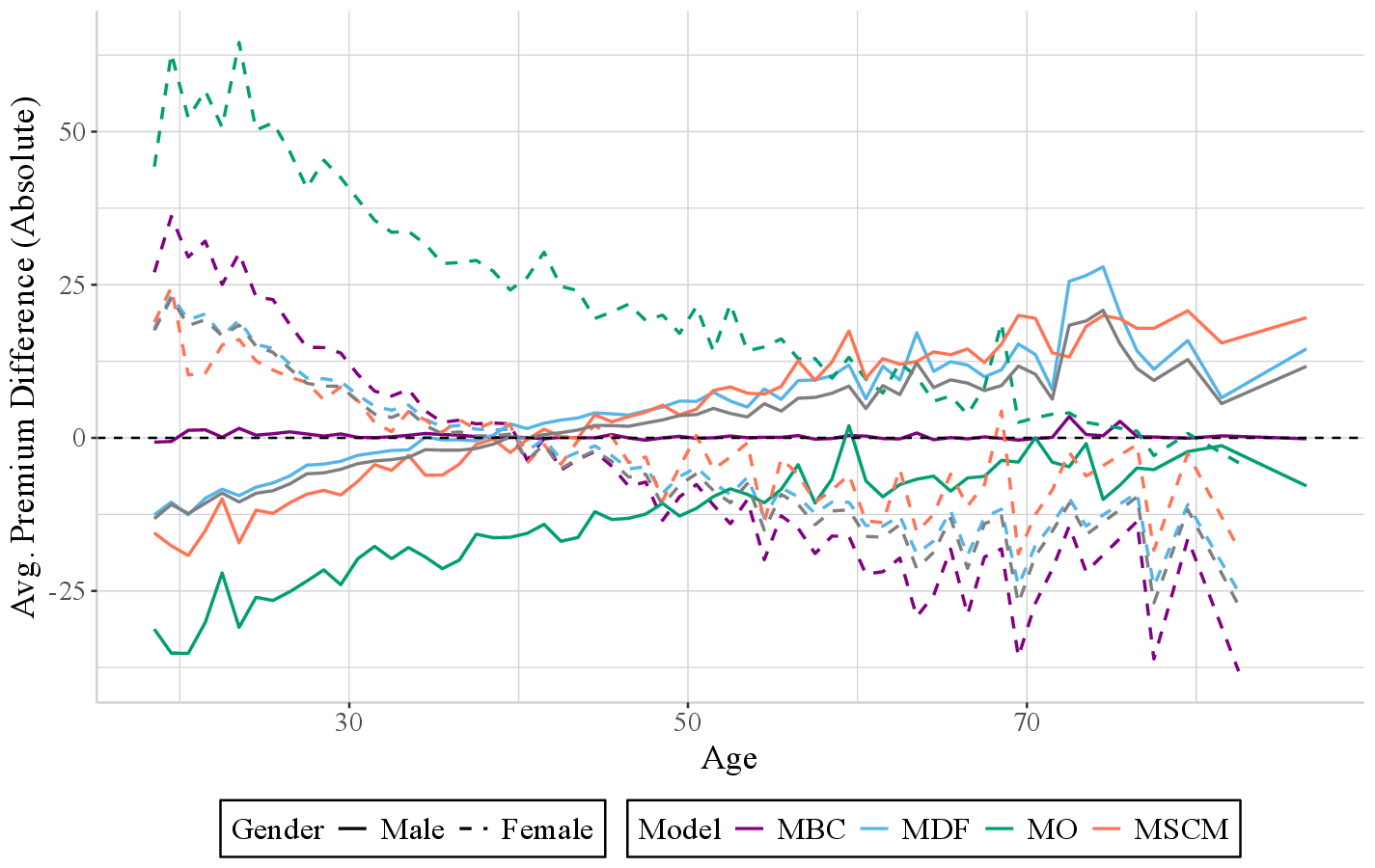}
    \caption{Absolute premium difference in \textit{fremotor1prem0304a} (GLM models versus GLM MB)}
    \label{fresol}
\end{figure}

As shown in Figure~\ref{pgsol}, group fairness is achieved via premium redistribution in the \textit{pg15training} dataset. This leads to a situation where lower-risk young female policyholders pay above their actuarially fair share under the Best Estimate (MB) model and young males pay less. This redistribution becomes excessive in fairness-adjusted models such as MU, MDF, MBC, and MSCM and the original risk differential is reversed. This suggests that the higher concentration of elderly females in the female cohort increases the group’s average risk, which leads younger and lower-risk females to cross-subsidize both their higher-risk peers and lower-risk males.

In contrast, cross-gender subsidization is less prevalent in the \textit{fremotor1prem0304a} dataset as shown in Figure~\ref{fresol}. Only the MO model exhibits a clear subsidy from young females to young males. In other models, the primary flow of cross-subsidization occurs within gender groups and younger policyholders subsidize older ones. Furthermore, the overall gender-based premium gap persists across most models, which suggests that age effects dominate the fairness adjustments in this dataset and the mechanisms for achieving group fairness do not uniformly induce gender-level redistribution. More plots regarding solidarity can be found in Appendix~\ref{soll}

\subsubsection{Adverse Selection}\label{sec:as}
Another critical concern in insurance is adverse selection, which arises from information asymmetry between insurers and policyholders. Fairness-driven pricing models may inadvertently alter premium structures in ways that attract higher-risk individuals or deter lower-risk ones, particularly if risk heterogeneity is not fully captured. To assess this risk, we employ the double lift chart methodology of \citet{goldburd2016generalized}, which evaluates the impact of shifting from a benchmark model to a fair model by sorting policyholders according to the ratio $\frac{Y_{\text{benchmark}}}{Y_{\text{fair}}}$. 

We analyze the actual claim amounts across deciles of this ratio to identify which groups are most encouraged (top bins, where fair premiums are lower) or discouraged (bottom bins, where fair premiums are higher). As discussed in Section~\ref{FTU}, directly removing the sensitive attribute is the most straightforward fairness approach. Hence, we use the Unawareness model (MU) as our benchmark and compare it to more sophisticated fair models—MO, MDF, and MSCM—to examine how different fairness mechanisms influence the risk profile of the attracted and retained customer base.

\begin{figure}[htbp]
\centering 
\subfigure{
\label{as:1}
\includegraphics[width=0.485\linewidth]{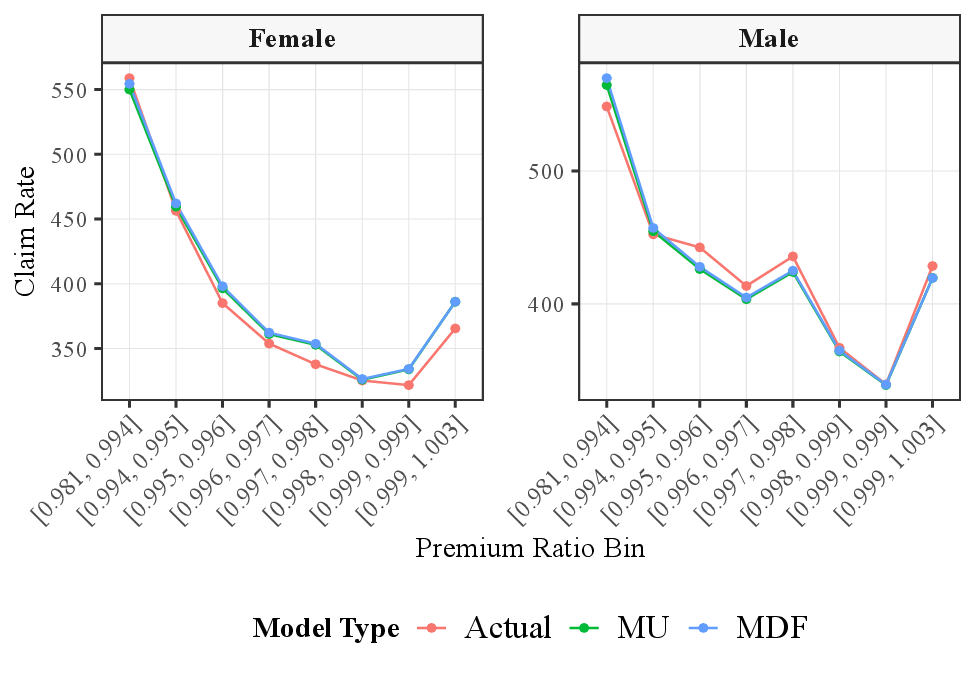}}
\subfigure{
\label{fre2}
\includegraphics[width=0.485\linewidth]{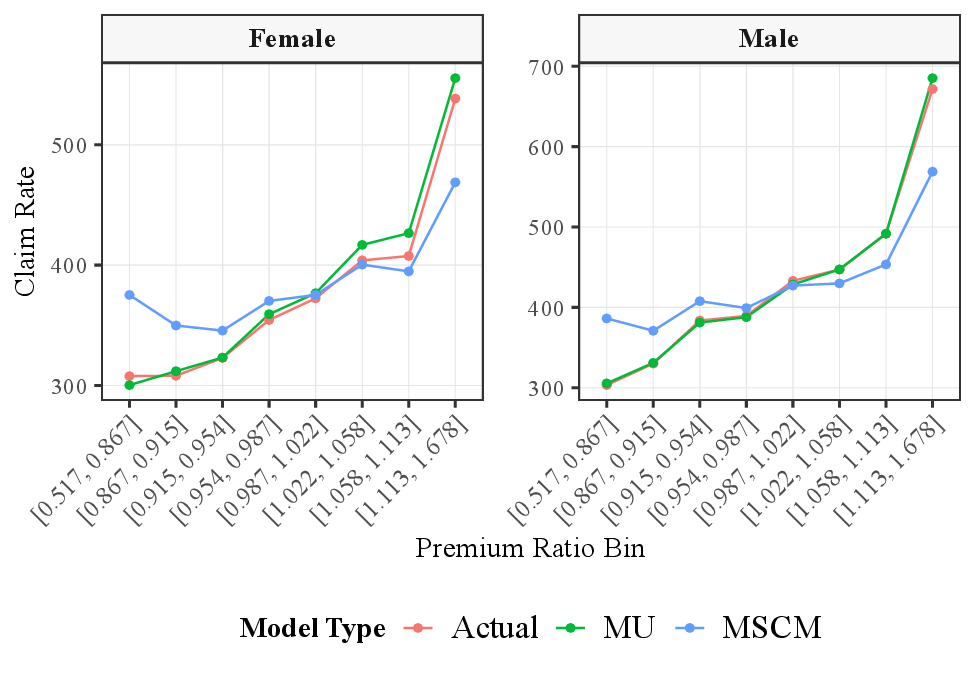}}
\caption{Double lift charts by gender (Left: GLM MDF versus GLM MU; Right: XGBoost MSCM versus XGBoost MU) in \textit{fremotor1prem0304a}}
\label{as fre}
\end{figure}

\begin{figure}[htbp]
\centering 
\subfigure{
\label{as:2}
\includegraphics[width=0.487\linewidth]{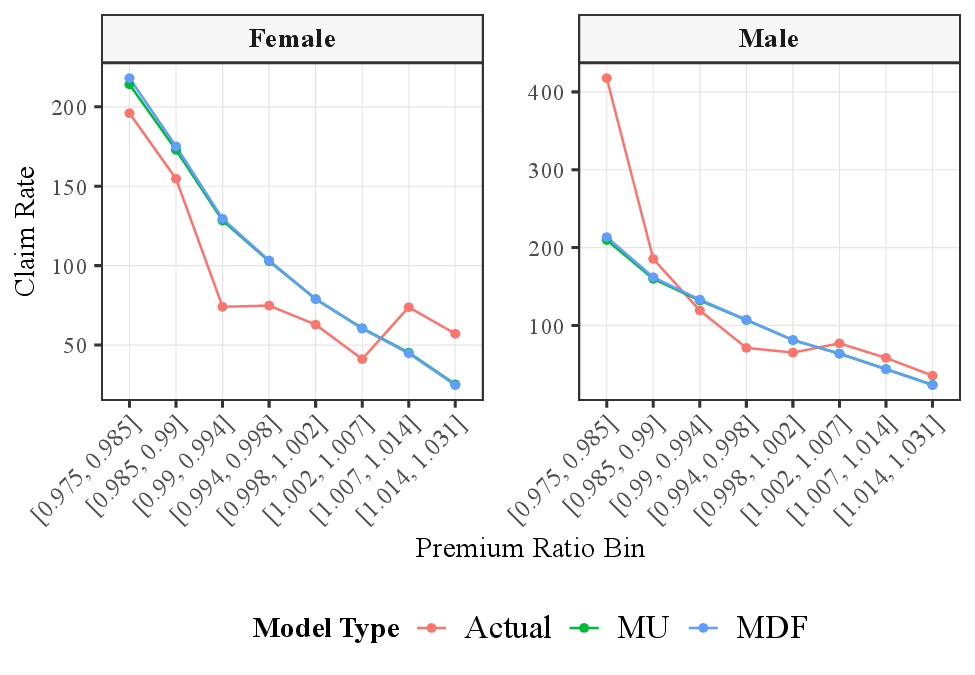}}
\subfigure{
\label{fre3}
\includegraphics[width=0.487\linewidth]{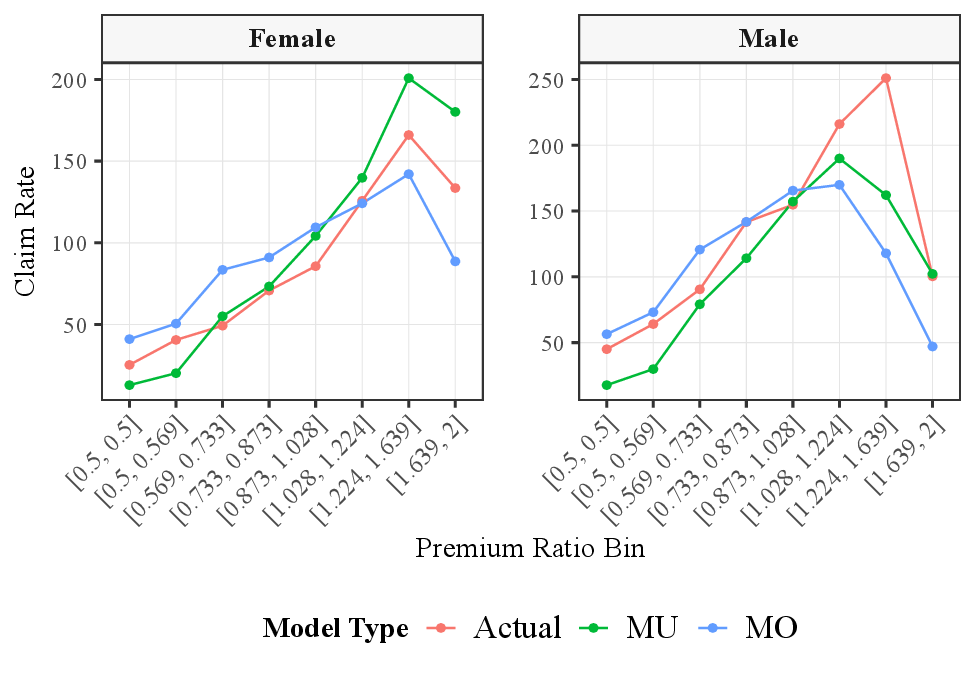}}
\caption{Double lift charts by gender (Left: GLM MDF versus GLM MU; Right: XGBoost MO versus XGBoost MU) in \textit{pg15training}}
\label{as pg}
\end{figure}

A consistent pattern across both datasets is that the GLM-based models tend to attract lower-risk customers while discouraging higher-risk ones—a favorable outcome for insurer profitability. In contrast, XGBoost exhibits a tendency to attract higher-risk individuals despite its higher predictive accuracy. This increases the risk of adverse selection and reveals a critical trade-off in insurance pricing model design: while complex machine learning models like XGBoost offer improved accuracy, they may inadvertently lead to undesirable market dynamics like adverse selection and exhibit poorer individual fairness, as discussed in Section~\ref{if}. Additional double lift charts are provided in Appendix~\ref{as}.

\subsection{Model Ensembling}

As suggested in Section~\ref{comparison}, multiple fairness notions often conflict with each other and with predictive accuracy. This inherent trade-off calls for a strategy that can balance competing objectives without compromising deployability, interpretability, or regulatory compliance.

\subsubsection{Meta Learner}

A natural multi-objective optimization that jointly optimizes accuracy and fairness is impractical. This is because fairness metrics such as counterfactual fairness are inherently complex and often require global computations over all predictions. Their non-differentiable and computationally-expensive nature makes joint optimization with accuracy unstable or intractable.

To address these challenges, we adopt a two-stage ensemble framework that separates fairness modeling from performance refinement. We use two specialized base learners: MO that is strong in group fairness, and MSCM that is effective for individual and counterfactual fairness. Both use sensitive attributes (e.g., gender) only during preprocessing and produce gender-free predictions at deployment. This satisfies regulatory and ethical requirements as well as ensuring that our model makes decisions without knowing the gender variable.

We deliberately omit the benchmark model that includes gender in final pricing (MB) despite its higher raw accuracy. This is because dependence on gender at the point of decision compromises explainability, accountability, and legal defensibility. Furthermore, while predictive performance can often be recovered through ensembling or post-hoc calibration, fairness violations embedded in model outputs are extremely difficult to mitigate after the fact.

In the second stage, we do not combine MO and MSCM by simple averaging that presumes uniform weighting across all risks. We make the combination of the two models through a neural network–based meta-learner. This enables context-sensitive weighting, which means that the ensemble may rely more heavily on MSCM when pricing low-frequency heterogeneous risks and place greater weight on MO for large homogeneous groups. This is because individual fairness is dominant for MSCM and maintaining group-level parity is the primary concern of MO. In this way, the framework automatically learns the pattern and emulates the adaptive judgment that experienced actuaries may also apply across different risk classes.

\subsubsection{NSGA-II}
As demonstrated in the model comparison in Section~\ref{comparison}, different fairness criteria are often compatible. This means that improving fairness typically comes at the cost of predictive accuracy. This trade-off transforms the task of model selection into a multi-objective optimization (MOO) problem, where the goal is to identify solutions that balance performance across competing objectives. MOO is a well-established field in mathematics, engineering, and decision theory. It has a wide range of methods developed for navigating such trade-offs \citep{miettinen1999nonlinear}. Gradient-based approaches are commonly employed due to their efficiency and convergence properties. These methods typically scalarize objectives into a single loss using techniques such as weighted sums or Tchebycheff norms \citep{miettinen1999nonlinear}. 
However, the choice of weights and normalization schemes is non-trivial and can significantly influence the outcome. The optimization may fail if the front is non-convex. Recent developments in differentiable MOO include Pareto navigation and first-order methods that compute descent directions by balancing gradients across objectives. A prominent example is Multiple Gradient Descent Algorithm (MGDA) \citep{desideri2012multiple}, which finds a common descent direction by solving a quadratic subproblem. Despite their appeal, such gradient-based methods require all objectives to be differentiable. This poses a major challenge in fair machine learning, where key fairness metrics (such as demographic parity gap or local Lipschitz constant) are often non-differentiable and discontinuous.

\begin{algorithm}[htbp]
\caption{NSGA-II: Non-dominated Sorting Genetic Algorithm II}
\small
\label{alg:nagasii}
\begin{algorithmic}
\REQUIRE Population size $N$, Maximum generations $G_{max}$, Crossover probability $p_c$, Mutation probability $p_m$
\ENSURE Final non-dominated set $\mathcal{P}^*$

\STATE Generation counter: $t \gets 0$
\STATE Initialize random population $P_0$ of size $N$
\STATE Evaluate objectives for each individual in $P_0$
\STATE Fast Non-dominated-Sort($P_0$) \COMMENT{Assign rank and crowding distance}

\FOR{$t = 1$ to $G_{max}$}
    \STATE $Q_t \gets \text{Make-New-Population}(P_t)$ \COMMENT{Selection, crossover, mutation}
    \STATE Evaluate objectives for each individual in $Q_t$
    \STATE $R_t \gets P_t \cup Q_t$ \COMMENT{Combine parent and offspring populations}
    \STATE $\mathcal{F} \gets \text{Fast-Non-dominated-Sort}(R_t)$ \COMMENT{$\mathcal{F} = (F_1, F_2, \dots)$}
    
    \STATE $P_{t+1} \gets \emptyset$
    \STATE $i \gets 1$
    \WHILE{$|P_{t+1}| + |F_i| \leq N$} 
        \STATE Calculate-Crowding-Distance($F_i$) \\
        \STATE $P_{t+1} \gets P_{t+1} \cup F_i$\\
        \STATE $i \gets i + 1$\\
    \ENDWHILE
    
    \STATE Sort($F_i$, $>_n$), where $>_n$ is a standard crowded-comparison operator \\
    \STATE $P_{t+1} \gets P_{t+1} \cup F_i[1:(N-|P_{t+1}|)]$ \\
    \STATE $t \gets t + 1$
\ENDFOR
\RETURN Non-dominated solutions from $P_{G_{max}}$
\end{algorithmic}
\end{algorithm}
To overcome this limitation, derivative-free methods like Evolutionary Algorithms offer a robust alternative. Among them, Non-dominated Sorting Genetic Algorithm II (NSGA-II) is a benchmark MOO algorithm that does not rely on gradients or convexity assumptions. As demonstrated by \cite{quadrianto2017recycling} and \cite{robertson2024human}, NSGA-II has been successfully applied in fair machine learning contexts. This makes it a suitable choice for optimizing models under multiple non-differentiable fairness constraints. NSGA-II operates on the foundational principles of Darwinian evolution—selection, crossover, and mutation. NSGA-II introduces two critical innovations that distinguish it from earlier MOO algorithms: non-dominated sorting and crowding distance estimation \citep{deb2002fast}. The algorithm proceeds iteratively through a sequence of generations and each generation comprises four core steps (see Algorithm~\ref{alg:nagasii}). First, an initial population is randomly generated and evaluated on all objectives. Solutions are then ranked via non-dominated sorting into hierarchical fronts, with the first front containing all non-dominated solutions. Within each front, crowding distance is computed as the average perimeter of the cuboid formed by nearest neighbors in the objective space to promote diversity. A new population is created through binary tournament selection that favors lower rank and higher crowding distance, followed by crossover and mutation. The combined parent and offspring population of size $2N$ undergoes another round of non-dominated sorting and crowding distance assignment. The best $N$ solutions are selected for the next generation by prioritizing dominance rank first and diversity second. For further details on NSGA-II’s theoretical properties, computational efficiency, and suitability in fairness-aware optimization, see Appendix~\ref{nsga}.

\subsubsection{Technique for Order of Preference by Similarity to Ideal Solution (TOPSIS)}
Following the generation of the Pareto-optimal front via NSGA-II, the selection of a single optimal solution from the set of non-dominated trade-offs remains a critical challenge in multi-objective decision making. This is particularly the case in the context of fair insurance pricing, where conflicting objectives such as profitability, risk equity, and regulatory compliance must be reconciled. While traditional scalarization methods (e.g., weighted sum, $\varepsilon$-constraint) often require convexity of the Pareto front and may fail to capture non-convex or disconnected regions \citep{miettinen1999nonlinear}, 
TOPSIS offers a robust, intuitive, and geometry-based alternative that does not rely on such restrictive assumptions. TOPSIS (see Algorithm~\ref{alg:topsis}), developed by \cite{hwang1981methods},  identifies the utopia and nadir solutions, which represent the best and worst possible outcomes across all criteria. It ranks alternatives according to their relative closeness to the utopia solution, computed from their normalized Euclidean distances to both the utopia and nadir points.

This method is particularly well-suited for integration with NSGA-II in insurance pricing applications because it preserves the diversity of the Pareto front while enabling a transparent and criterion-weighted selection process that accommodates stakeholder preferences. Crucially, the weight assignment in TOPSIS allows regulators, insurers, and actuaries to explicitly encode their priorities (e.g., emphasizing fairness over profit, or vice versa) through normalized criterion weights. This formalizes a participatory governance mechanism within the algorithmic decision pipeline \citep{zhang2025multi}. By bridging the computational power of evolutionary multi-objective optimization with the normative flexibility of multi-criteria decision analysis, TOPSIS facilitates not only technically optimal but also socially and ethically defensible pricing strategies in complex and high-stakes domains like insurance. For a comprehensive overview of TOPSIS and its implementation, see \cite{behzadian2012state}.

\begin{algorithm}[htbp]
\caption{TOPSIS: Technique for Order of Preference by Similarity to Ideal Solution}
\label{alg:topsis}
\small
\begin{algorithmic}
\REQUIRE Set of $n$ Pareto-optimal solutions $\mathcal{P}^* = \{x_1, x_2, \dots, x_n\}$, $m$ objectives $f_1, \dots, f_m$, weights $w_j \in [0,1]$ such that $\sum_{j=1}^m w_j = 1$
\ENSURE Ranked solutions with optimal solution $x^*$

\STATE Construct decision matrix $\mathbf{D} \in \mathbb{R}^{n \times m}$, where $d_{ij} = f_j(x_i)$
\STATE Normalize decision matrix to obtain $\mathbf{R} = [r_{ij}]_{n \times m}$:
    \STATE \quad $r_{ij} = \dfrac{d_{ij}}{\sqrt{\sum_{k=1}^n d_{kj}^2}}$ 
\STATE Compute weighted normalized matrix $\mathbf{V} = [v_{ij}]_{n \times m}$:
    \STATE \quad $v_{ij} = w_j \cdot r_{ij}$
\STATE Determine ideal solution $A^+$ and anti-ideal solution $A^-$:
    \STATE \quad $A^+ = (v_1^+, v_2^+, \dots, v_m^+)$, where $v_j^+ = \begin{cases}
        \max\limits_{i} v_{ij} & \text{if } f_j \text{ is benefit criterion} \\
        \min\limits_{i} v_{ij} & \text{if } f_j \text{ is cost criterion}
    \end{cases}$\\
    \STATE \quad $A^- = (v_1^-, v_2^-, \dots, v_m^-)$, where $v_j^- = \begin{cases}
        \min\limits_{i} v_{ij} & \text{if } f_j \text{ is benefit criterion} \\
        \max\limits_{i} v_{ij} & \text{if } f_j \text{ is cost criterion}
    \end{cases}$
\STATE Compute separation measures for each solution $x_i$:\\
    \STATE  \quad Distance to ideal solution:$D_i^+ = \sqrt{\sum_{j=1}^m (v_{ij} - v_j^+)^2}$ \\
    \STATE \quad Distance to anti-ideal solution: $D_i^- = \sqrt{\sum_{j=1}^m (v_{ij} - v_j^-)^2}$ 
\STATE Compute relative closeness coefficient for each solution $x_i$:
    \STATE \quad $C_i = \dfrac{D_i^-}{D_i^+ + D_i^-}, \quad 0 \leq C_i \leq 1$
\STATE Rank solutions in descending order of $C_i$ values
\RETURN Optimal compromise solution: $x^* = \arg\max\limits_{i} C_i$ 
\end{algorithmic}
\end{algorithm}

\subsubsection{Implementation}
We apply NSGA-II to train our neural network meta learner. In this framework, the weights of the neural network serve as the decision variables in the evolutionary algorithm and form the population of potential solutions. Through non-dominated sorting and crowding distance mechanisms, the algorithm identifies a diverse set of Pareto-optimal models. To ensure the effectiveness of the evolutionary search, we conducted thorough hyperparameter tuning for NSGA-II. The hyperparameters include population size, crossover probabilities mutation probabilities, and the number of generations. The selection of final parameter values was guided by two primary criteria: (1) the quality of the resulting Pareto front, evaluated using hypervolume (a metric capturing both convergence and diversity), and (2) computational efficiency, which is critically dependent on dataset size and runtime constraints. This tuning is performed through a series of pilot runs on a validation set, aimed at stable convergence to a well-distributed Pareto front. 
\begin{table}[H]
\resizebox{\textwidth}{!}{%
\begin{tabular}{|c|c|c|c|c|c|}
\hline
Dataset                     & Initialization mode & Population size & Maximum number of generations & Crossover rate & Mutation rate \\
\hline
\textit{pg15training}       & Random              & 50              & 25                            & 90\%           & 10\%          \\
\hline
\textit{fremotor1prem0304a} & Random              & 120             & 50                            & 90\%           & 20\%         \\
\hline
\end{tabular}}
\label{tuning}
\caption{NSGA-II hyperparameter settings in two datasets}
\end{table}

After obtaining the Pareto front (see Appendix~\ref{pareto}), we need to select the optimal point with preference for different dimensions. We apply TOPSIS to select the solution closest to the ideal point and farthest from the anti-ideal point. In our experiment, the weights are given by $[0.3,0.3,0.3,0.1]$ for accuracy, group fairness, individual fairness, and  counterfactual fairness respectively.

\subsubsection{Ensembling Results}
\begin{figure}[htbp]
\centering 
\subfigure{
\label{1:1}
\includegraphics[width=0.44\linewidth]{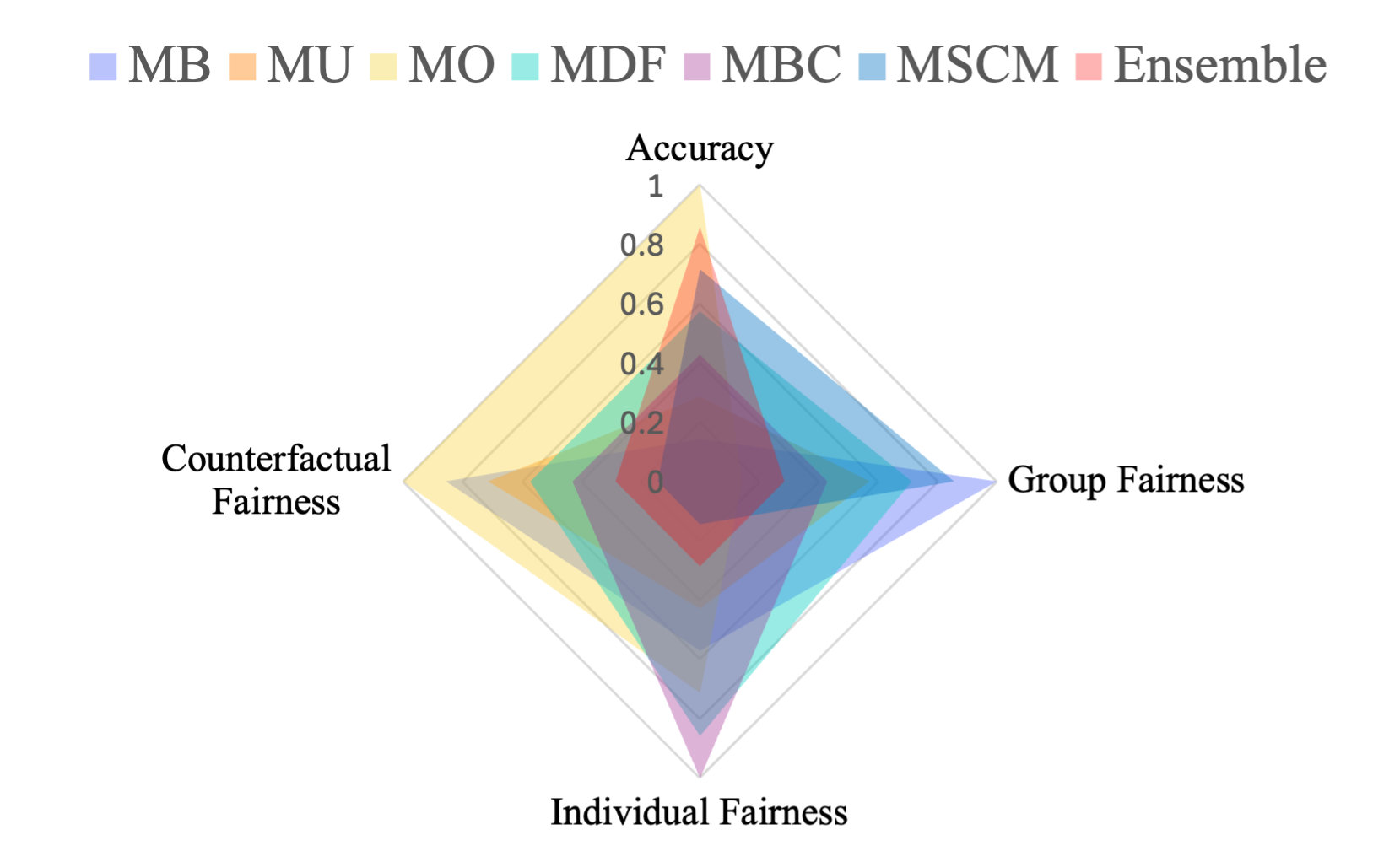}}
\subfigure{
\label{fre4}
\includegraphics[width=0.44\linewidth]{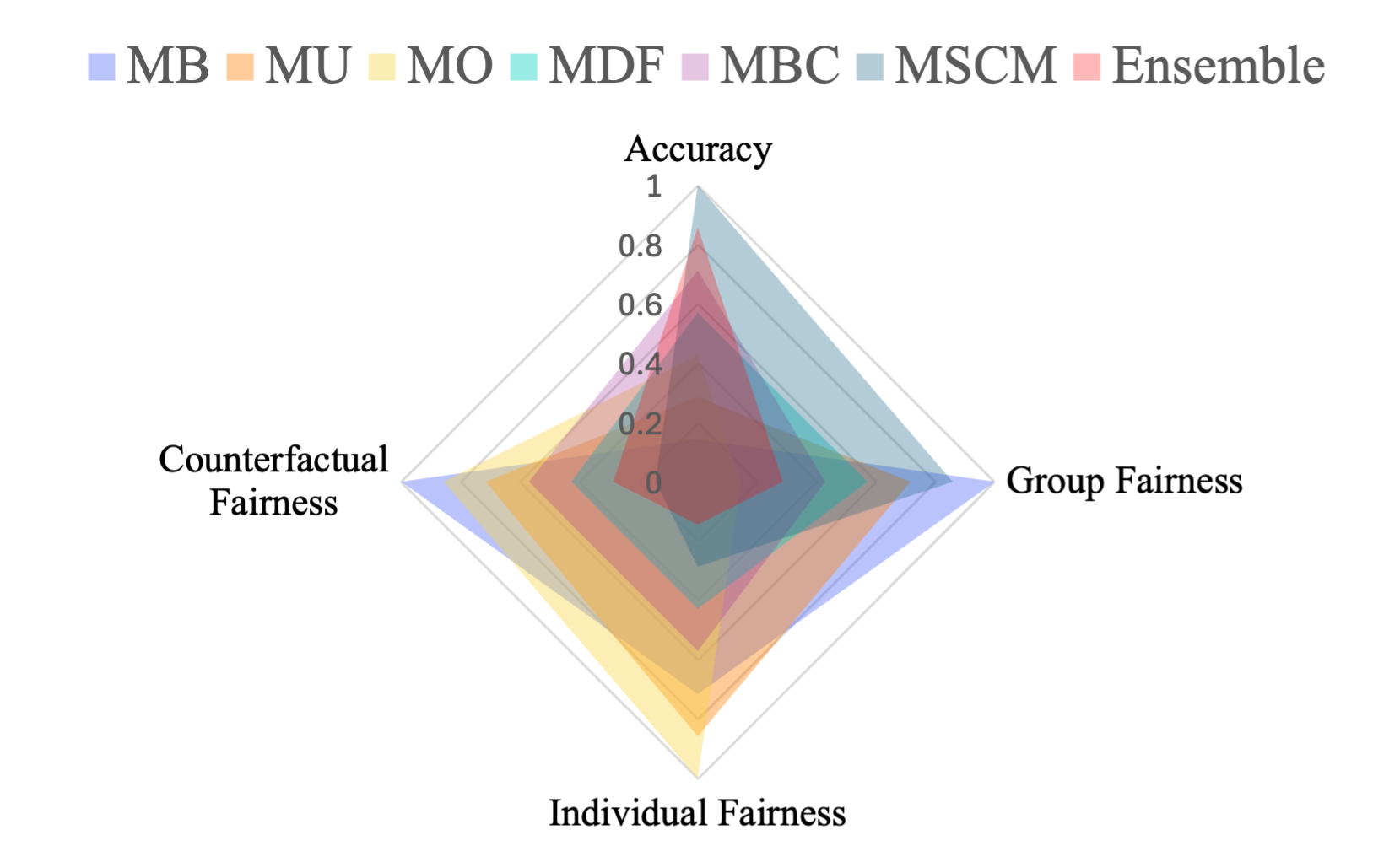}}
\caption{Comparison of performances in \textit{fremotor1prem0304a}(Left:GLM, Right: XGBoost)}
\label{p2}
\end{figure}

\begin{figure}[htbp]
\centering 
\subfigure{
\label{1:2}
\includegraphics[width=0.44\linewidth]{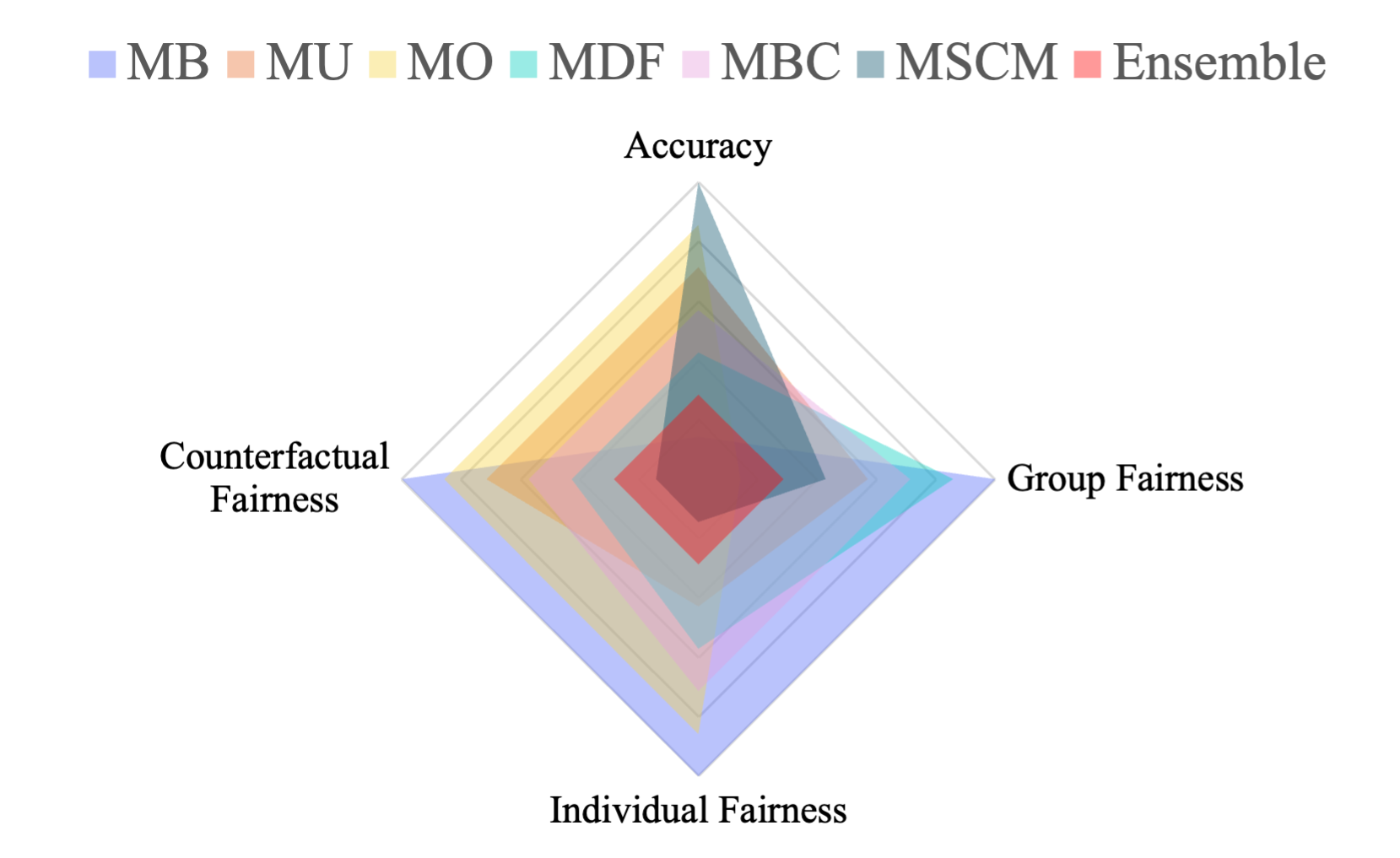}}
\subfigure{
\includegraphics[width=0.44\linewidth]{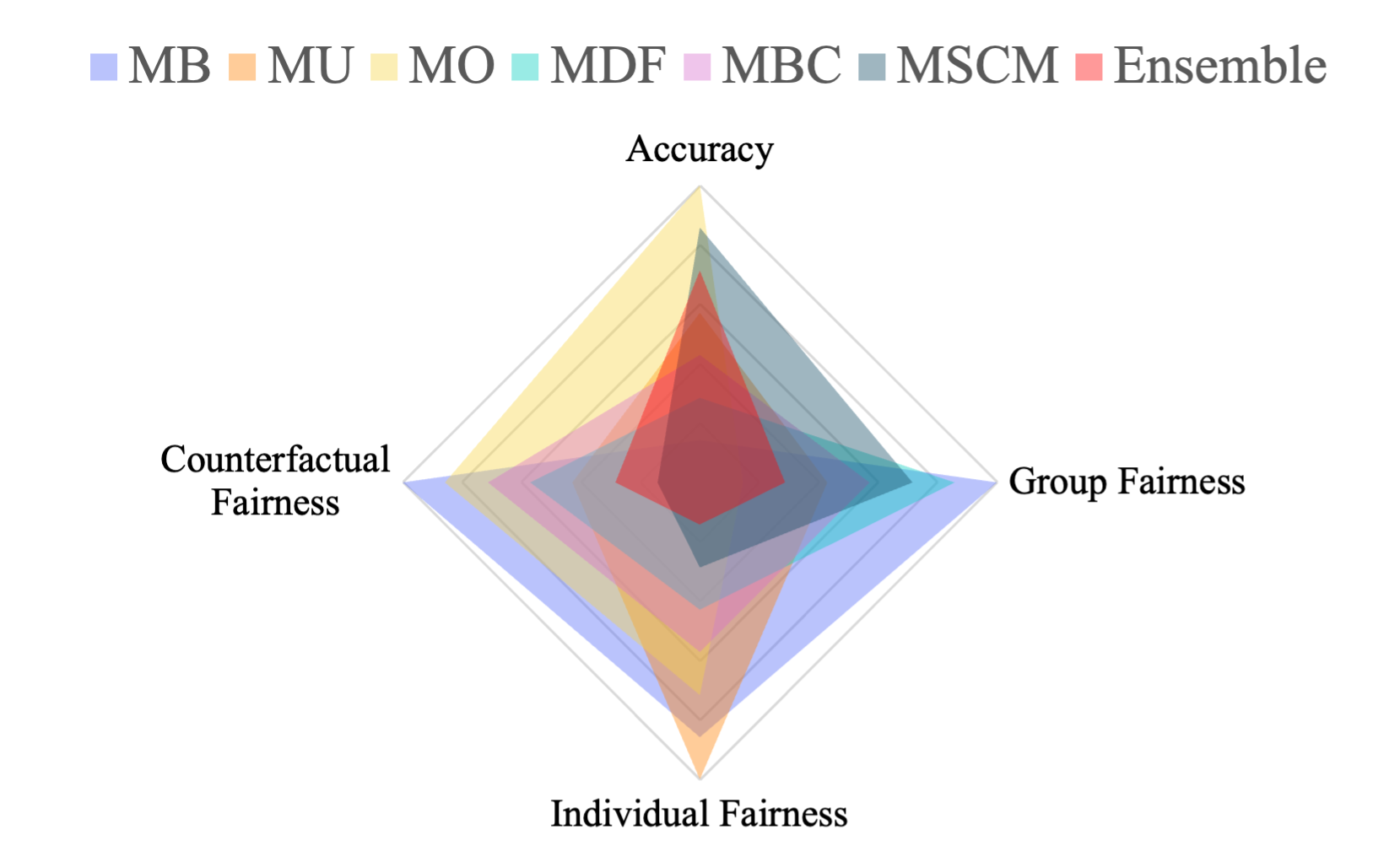}}
\caption{Comparison of performances in \textit{pg15training}(Left:GLM, Right: XGBoost)} 
\label{pg2}
\end{figure}

The specific metric comparison of the ensemble model with other single base learners is shown in Table~\ref{tab:model_comparison}, and the full results of the model ensembling are visualized using radar plots (see Figures~\ref{p2} and~\ref{pg2}). For comparability, each metric is standardized such that a lower value indicates better performance. To enhance visual clarity, we transform the scores into rank statistics within each dimension.

\begin{table}[H]
\centering
\scalebox{0.8}{
\begin{tabular}{|c|c|c|c|c|c|c|}
\hline
Model & Accuracy & Group Fairness & Individual Fairness & Counterfactual Fairness \\
\hline
MB & 459.10 &0.7426 & 1397.592 & -19.9600 \\
MU & 462.37 & 0.8915 & 1399.938 & 1.1428 \\
MO & 465.78 & 0.9462 & 1391.251 & -6.1700 \\
MDF & 461.30 &  0.8847 & 1362.751 & 1.2530 \\
MBC & 461.32 & 0.8888 & 1367.633 & 4.8200 \\
MSCM & 462.90 & 0.8857 & 1316.120 & 0.0954 \\
Ensemble & 462.71& 0.9355 & 1272.489 & -1.0350 \\
\hline
\end{tabular}}
\caption{Model performance comparison across different fairness metrics on the \textit{pg15training} dataset using XGBoost. The reported statistics include accuracy measured by RMSE (lower is better), group fairness quantified by the disparity ratio (closer to 1 is better), individual fairness (lower is better), and counterfactual fairness (closer to 0 is better).} 
\label{tab:model_comparison}
\end{table}
As shown in the plots, the ensemble model achieves a well-balanced performance across all criteria in both datasets. It outperforms MSCM in terms of accuracy and group fairness, while surpassing MO in individual and counterfactual fairness. This demonstrates the effectiveness of the NSGA-II-optimized neural meta-learner in leveraging the complementary strengths of its base models, resulting in a robust and fair predictive system.

\section{Conclusion}\label{sec:conclusion}
The use of machine learning in insurance pricing enhances accuracy but risks algorithmic bias. Moving beyond group fairness, we evaluate fairness-aware models on two real-world motor insurance datasets using a multidimensional fairness framework. We find that group, individual, and counterfactual fairness often conflict with each other and with accuracy—no single model excels in all dimensions. MO improves group fairness by removing proxy effects but harms counterfactual fairness due to feature distortion. MSCM preserves counterfactual and individual fairness better than MO by minimizing individual treatment effects. Both models improve fairness at the cost of lower accuracy.

To balance these trade-offs, we propose a multi-objective NSGA-II framework to generate a Pareto front of ensemble models optimizing accuracy and multiple fairness criteria. We use TOPSIS to select solutions aligned with regulatory or business priorities. Our final ensemble model achieves a balanced performance across all objectives and outperforms individual fairness-aware baselines in overall trade-off quality. Economically, enforcing demographic parity leads to a modest cross-subsidization effect. We also find that more accurate models may attract higher-risk customers, which exacerbates adverse selection. Together, our results provide a practical and ethically informed approach to fair and sustainable AI-driven insurance pricing.

Several promising directions for future research emerge. First, the analysis could be extended from pure premium estimation to the full insurance pricing pipeline. This includes demand modeling and price optimization under fairness or regulatory constraints \citep{shimao2025welfare}. Second, the framework may be adapted to other insurance lines, such as life insurance or annuities, since longevity risk and intergenerational equity pose distinct fairness challenges \citep{lau2024deferred}. Finally, future studies could address continuous variables, which entail unique challenges in defining and enforcing fairness \citep{grari2019fairness,lee2025machine}.
\appendix
\setlength{\bibsep}{0.0pt}
\bibliographystyle{apalike}
{\footnotesize\bibliography{ref}}

\section{Regulations}\label{reg}

In this section, we examine rate regulations related to price discrimination in insurance. We categorize these regulations according to the underlying discrimination or fairness criterion they address as defined in Section~\ref{def}. This classification reflects the diverse perspectives on fairness and different views of insurance.
\subsection{Regulations on Direct Discrimination}
The most direct form of rate regulation involves prohibiting the use of specific sensitive attributes in pricing models which  corresponds to the notion of FTU (see Section~\ref{FTU}). The definition of what constitutes a sensitive attribute varies significantly across jurisdictions and lines of insurance, reflecting differing legal and ethical standards. A prominent example is the European Union's Gender Directive (2004/113/EC), which mandates that gender must not be used as a factor in the calculation of insurance premiums or benefits, and establishes a principle of gender-neutral pricing. Similarly, in New South Wales, Australia, the \textit{Anti-Discrimination Act 1977} (Section 49Q) prohibits the use of disability as a pricing factor unless the insurer can demonstrate that the differential treatment is based on reliable actuarial or statistical data. These regulations exemplify a regulatory approach that prioritizes fairness and non-discrimination, requiring insurers to justify any risk classification based on protected characteristics with robust, objective evidence.

\subsection{Regulations on Indirect Discrimination}
Similar to regulations targeting direct discrimination, rules addressing indirect discrimination also limit the permissible scope of variables in insurance pricing. They focus specifically on proxy variables—factors that, while not sensitive themselves, strongly correlate with protected attributes. However, the use of such regulations often requires robust statistical evidence demonstrating the actuarial relevance of a variable, and showing that its exclusion would effectively mitigate indirect discrimination \citep{mosley2021methods}.
For instance, the U.S. Genetic Information Nondiscrimination Act (GINA) regulates the use of genetic information by broadly defining it to include not only individual genetic test results but also family medical history, participation in genetic research, and utilization of genetic counseling services \citep{feldman2012genetic}. This comprehensive definition aims to prevent insurers from using indirect indicators as proxies for genetic risk which is immutable and sensitive. Similarly, insurance regulators in Washington State have moved to restrict or ban the use of credit-based insurance scores, citing concerns that these scores may act as proxies for race and socioeconomic status. Ongoing investigations into potential civil rights violations underscore the growing regulatory scrutiny of seemingly neutral variables that perpetuate historical inequities, even if they exhibit predictive power.

\subsection{Regulations on Group Fairness}
Regulations grounded in group fairness often emphasize the principle of demographic parity, which requires that individuals from different protected groups receive similar average premiums. A prominent legal framework embodying this criterion is the doctrine of disparate impact. Under Title VII of the 1964 Civil Rights Act, employers are prohibited from using facially neutral practices that result in unjustified adverse effects on members of protected groups. A facially neutral practice appears non-discriminatory on its surface but leads to discriminatory outcomes in application. This principle was operationalized through the "80\% rule" (or four-fifths rule), first introduced in the 1972 California Guidelines on Employee Selection Procedures and later codified in the 1978 Uniform Guidelines on Employment Selection Procedures by the U.S. Equal Employment Opportunity Commission (EEOC) \citep{biddle2017adverse}. The rule presumes adverse impact if the selection rate for a protected group is less than 80\% of the rate for the most favored group. Similar protections are extended by the Age Discrimination in Employment Act of 1967 and the Fair Housing Act of 1968, which recognize disparate impact as a valid cause of action. However, its direct applicability to insurance pricing has been debated, particularly due to the actuarial reliance on risk segmentation \citep{xin2024antidiscrimination}.

An alternative, more extreme approach for achieving group fairness is community rating, which mandates uniform premiums for all individuals purchasing the same insurance product with guarantee, regardless of individual risk characteristics. This model fully embraces risk pooling and eliminates any form of risk-based differentiation. In Australia, private health insurance has operated under community rating since the enactment of the National Health Act 1953 and reinforced by the Private Health Insurance Act 2007. The premiums are set without regard to health status, age, claims history, or pre-existing medical conditions—key underwriting factors in traditional insurance models. Pure community rating permits premium variation only on the basis of benefit design and family composition, as implemented in countries such as the Netherlands and Switzerland \citep{thomson2009private}. In contrast, the United States employs an adjusted community rating system, which allows insurers to vary premiums based on certain demographic factors, including age and tobacco use \citep{trish2018does}.

\subsection{Regulations on Individual Fairness}
The concept of Individual Fairness, as defined in Section~\ref{fta}, aligns closely with the insurance regulatory principle of "unfair discrimination." This term is grounded in actuarial fairness, particularly as articulated in Principle 4 of the Casualty Actuarial Society’s (CAS) Statement of Principles Regarding Property and Casualty Insurance Ratemaking, which states that a rate is reasonable and not unfairly discriminatory if it constitutes an actuarially sound estimate of the expected future costs associated with an individual risk. This principle implies that individuals with similar risk profiles should be charged similar premiums. It reflects the broader statutory requirement—common across jurisdictions—that insurance rates must not be excessive, inadequate, or unfairly discriminatory \citep{chibanda2022defining}. Under this framework, unfairness arises not from differences in risk, but from differential treatment of comparable risks. For instance, Texas Insurance Code § 544.0002 explicitly prohibits insurers from charging an individual a different rate from rates charged to other individuals for the same coverage due to the individual's race, color, religion, or national origin. 

Furthermore, regulations on price optimization are also relevant to this discussion. Price optimization techniques, which adjust premiums based on consumer behavior, risk aversion, or price sensitivity rather than risk, can lead to differential pricing among individuals with similar risk exposure. This practice was deemed a form of unfair discrimination by the state of Maryland, which became the first U.S. state to ban price optimization in all lines of property and casualty insurance as of October 31, 2014 \citep{minty2016price}. After Maryland, seventeen states including California and Pennsylvania have joined in the same banning action.

\subsection{Regulations on Counterfactual Fairness}
With counterfactual fairness being a relatively new concept, there are currently no major, explicit regulations that formally require insurers to prove causal justification for risk factors. Instead, the bar remains "actuarial soundness" based on statistical correlation and predictive power. Moving to a causal standard would be a significant shift, requiring new methodologies for proving causality in complex real-world data and would likely face substantial industry resistance.
\section{Dataset Description}\label{dataset}
\begin{table}[H]
\caption{Dataset description for \textit{pg15training}}
\resizebox{\textwidth}{!}{%
\begin{tabular}{|l|l|}
\hline
\emph{Non-discriminatory Variable} & Description                                                                                           \\
\hline\hline
Bonus    & The bonus-malus (French no-claim discount): \\ & negative means bonus while positive means malus           \\
\hline
Group1   & The group of the car                                                                                  \\
\hline
Density  & The density of inhabitants (number of inhabitants per km2)\\ & in the city the driver of the car lives in \\
\hline
Value    & The car value (in euro)    \\    
\hline
\end{tabular}}
\label{tpg}
\end{table}

\begin{table}[H]
\caption{Dataset description for \textit{fremotor1prem0304a}}
\resizebox{\textwidth}{!}{%
\begin{tabular}{|l|l|}
\hline
\emph{Non-discriminatory Variable} & Description                                                                                           \\
\hline
\hline
BonusMalus    & Bonus/malus, between 50 and 350: $<$100 means bonus, $>$100 means malus in France           \\
\hline
PayFreq   & The payment frequency (as factor)                                                                               \\
\hline
JobCode  & The job code (as factor) \\
\hline
VehAge    & The vehicle age, in years   \\    
\hline
VehClass & The vehicle class (as factor) \\ 
\hline\hline
\emph{Risk factor} & Description     \\                   \hline                                             
VehPower&The vehicle power (as factor) from least powerful P2 to most powerful car P15\\
\hline
VehGas&The car gas, Diesel or regular (as factor)\\
\hline
VehUsage&The vehicle usage (as factor)\\
\hline
Garage& The type of garage (as factor)\\
\hline
Area&The area code (as factor)\\
\hline
Region&The policy regions in France (based on a standard French classification)\\
\hline
Channel&The channel distribution code (as factor)\\
\hline
\end{tabular}}
\label{tfre}
\end{table}

\section{Pareto Front}\label{pareto}
The Pareto front is visualized using a Parallel Coordinates Plot, in which each line corresponds to a distinct solution, and its performance across four evaluation metrics is displayed along the horizontal axes.\footnote{All metrics are aligned to a minimization objective: lower RMSE indicates higher predictive accuracy; the three fairness criteria, including Disparity Impact Ratio (group), Local Lipschitz Constant (individual), and Median ITE (counterfactual), are standardized or transformed such that smaller values correspond to greater fairness.} Within each plot, the solution that achieves the optimal (i.e., minimal) value for a specific metric is highlighted, thereby facilitating a clear and intuitive assessment of the trade-offs among the competing objectives.

\begin{figure}[h!]
    \centering
    \includegraphics[width=0.7\linewidth]{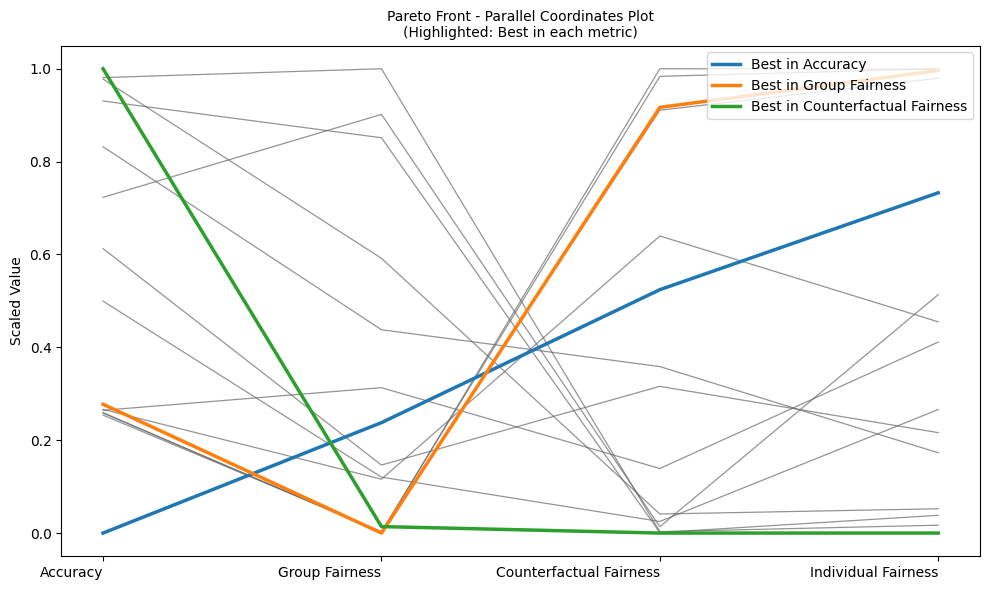}
    \caption{Pareto front for XGBoost}
    \label{1}
\end{figure}

\begin{figure}[h!]
    \centering
    \includegraphics[width=0.7\linewidth]{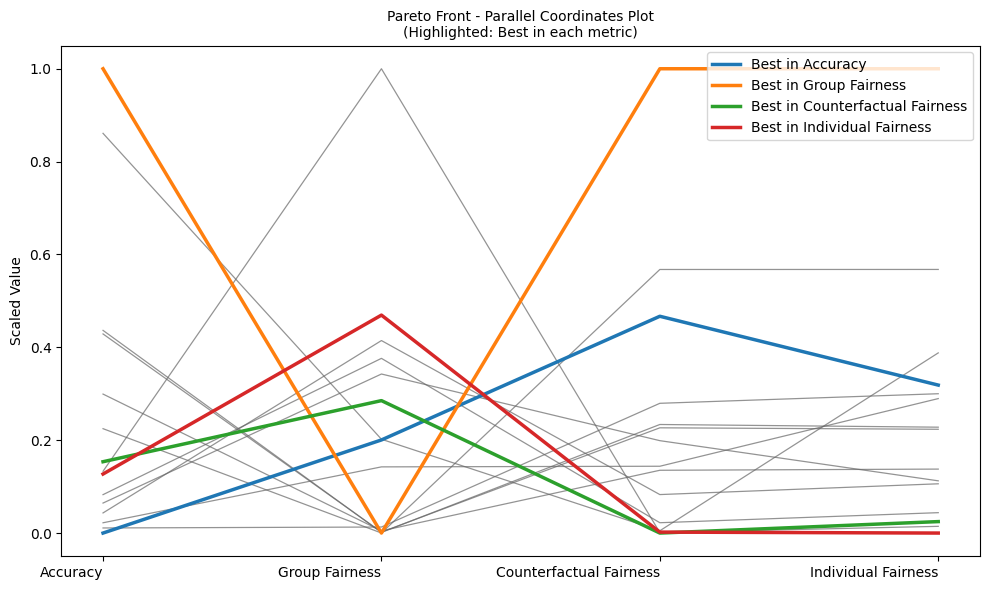}
    \caption{Pareto front for GLM}
    \label{2}
\end{figure}

\begin{figure}[h!]
    \centering
    \includegraphics[width=0.7\linewidth]{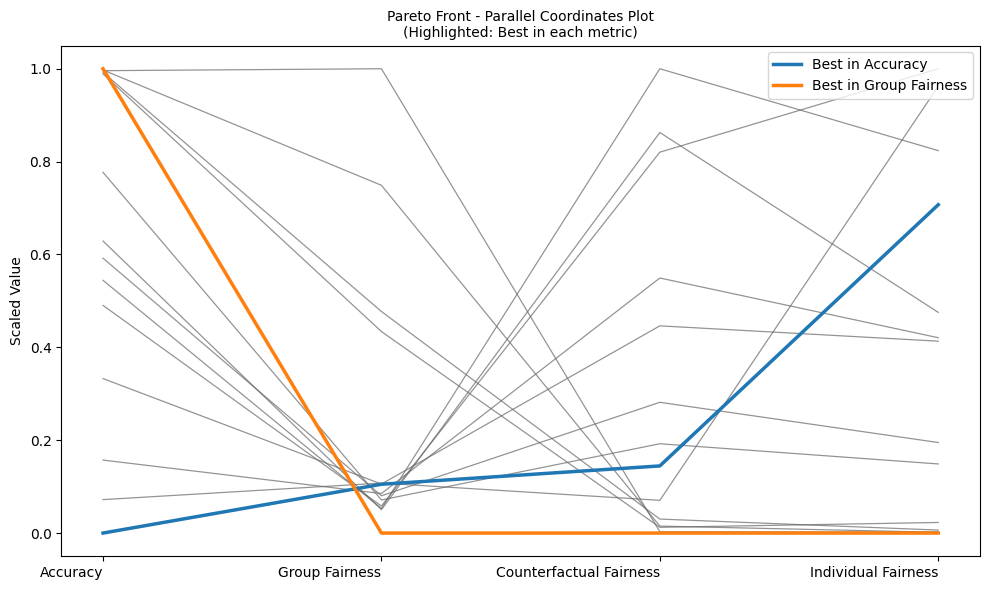}
    \caption{Pareto front for XGBoost}
    \label{3}
\end{figure}

\begin{figure}[h!]
    \centering
    \includegraphics[width=0.7\linewidth]{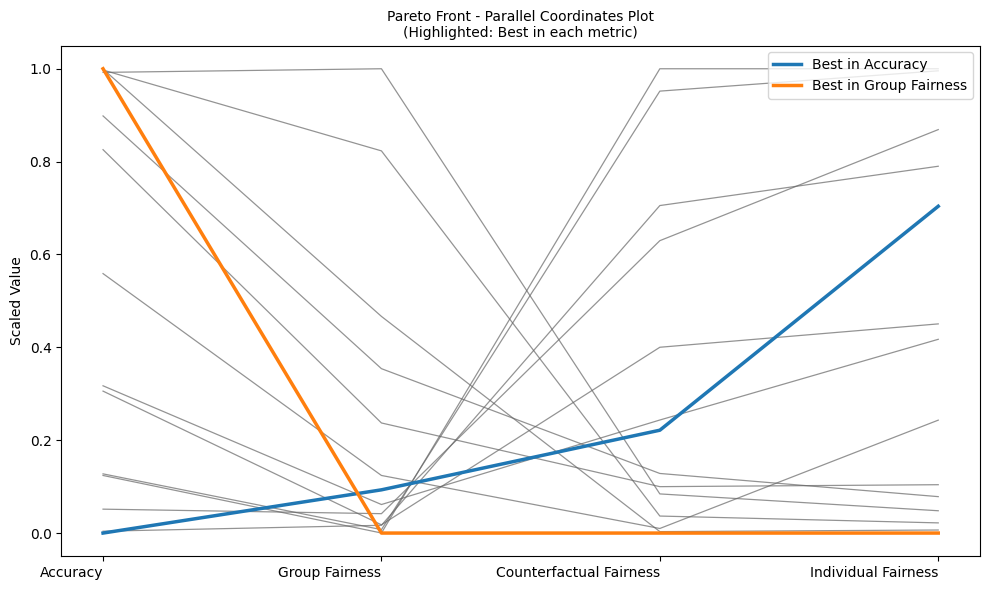}
    \caption{Pareto front for GLM}
    \label{4}
\end{figure}

For dataset \textit{pg15training}, the Pareto fronts are shown in Figure~\ref{1} and Figure~\ref{2}. 
For dataset \textit{fremotor1prem0304a}, the Pareto fronts are shown in Figure~\ref{3} and Figure~\ref{4}.
From the Pareto front, we can see the trade-off clearly through the fluctuations of lines. A key observation is that no solution dominates all others across all four metrics, confirming the inherent conflict between optimizing accuracy and enforcing fairness constraints. 

For the GLM-based meta-learner, the dominant trade-off is between group fairness and the other objectives, as evidenced by the pronounced “V”-shaped frontier concentrated along the group fairness axis (as shown in Figure~\ref{2}). Additionally, individual and counterfactual fairness metrics exhibit strong alignment across solutions, as most Pareto-optimal solutions show relatively flat trajectories in these two dimensions. This suggests that improvements in one tend to improve the other with minimal conflict.

In contrast, the XGBoost-based meta-learner exhibits a more global trade-off: accuracy versus all three fairness criteria simultaneously. Notably, its Pareto front consistently includes an extreme solution which achieves very poor accuracy but near-perfect scores across all fairness metrics (see the green solution in Figure~\ref{1} and orange solution in Figure~\ref{3}). This point strongly suggests a “collapsed” model—one that assigns the same prediction to every individual, trivially satisfying fairness definitions.

Why does this extreme solution emerge readily with XGBoost but less with GLM? The answer lies in the properties of their base predictions. XGBoost yields highly non-linear, dispersed outputs that vary significantly even across similar individuals. Under strong fairness constraints, the meta-model can most efficiently satisfy all fairness objectives by collapsing to a constant prediction, which achieves near-perfect fairness at the cost of accuracy and is quickly identified due to its Pareto dominance.

By contrast, GLM produces smoother, more structured predictions with inherently smaller disparities. The meta-model therefore tends to improve fairness via fine-grained adjustments without drastic accuracy loss, making the constant-output solution Pareto-inferior in most settings. That said, it remains reachable: as Figure~\ref{4} shows, when the dataset is smaller (with the same number of evolutionary generations), the GLM-based meta-model does converge to collapse. This confirms that GLM’s linearity raises—but does not eliminate—the barrier to collapse, and the solution can still emerge given sufficient exploration or reduced data complexity. 

\section{Additional Solidarity Plots}\label{soll}
This appendix provides supplementary solidarity plots that supplement our analysis in Section \ref{sec:sol}.
\begin{figure}[H]
    \centering
    \includegraphics[width=0.65\linewidth]{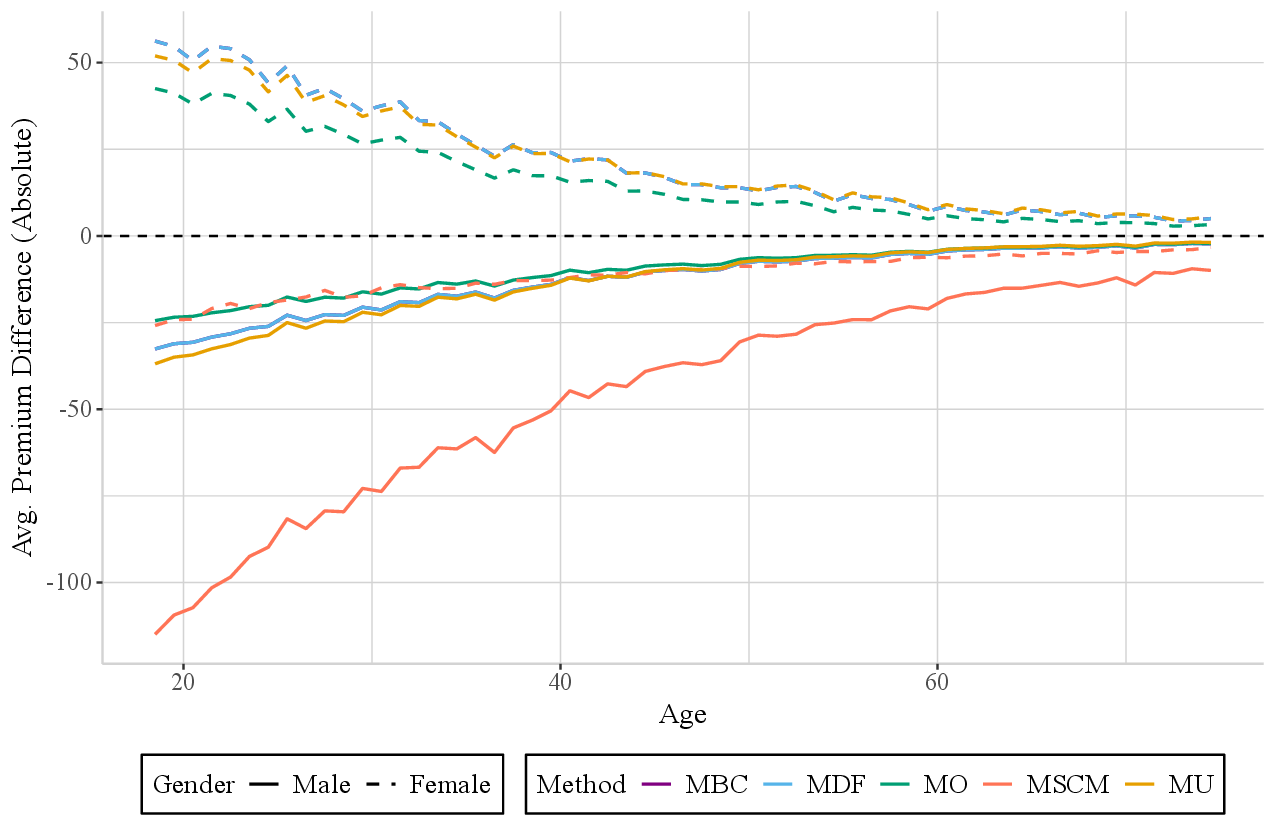}
    \caption{Relative premium difference in \textit{pg15training} (GLM models versus GLM MB)}
    \label{fig:placeholder}
\end{figure}
\begin{figure}[H]
    \centering
    \includegraphics[width=0.65\linewidth]{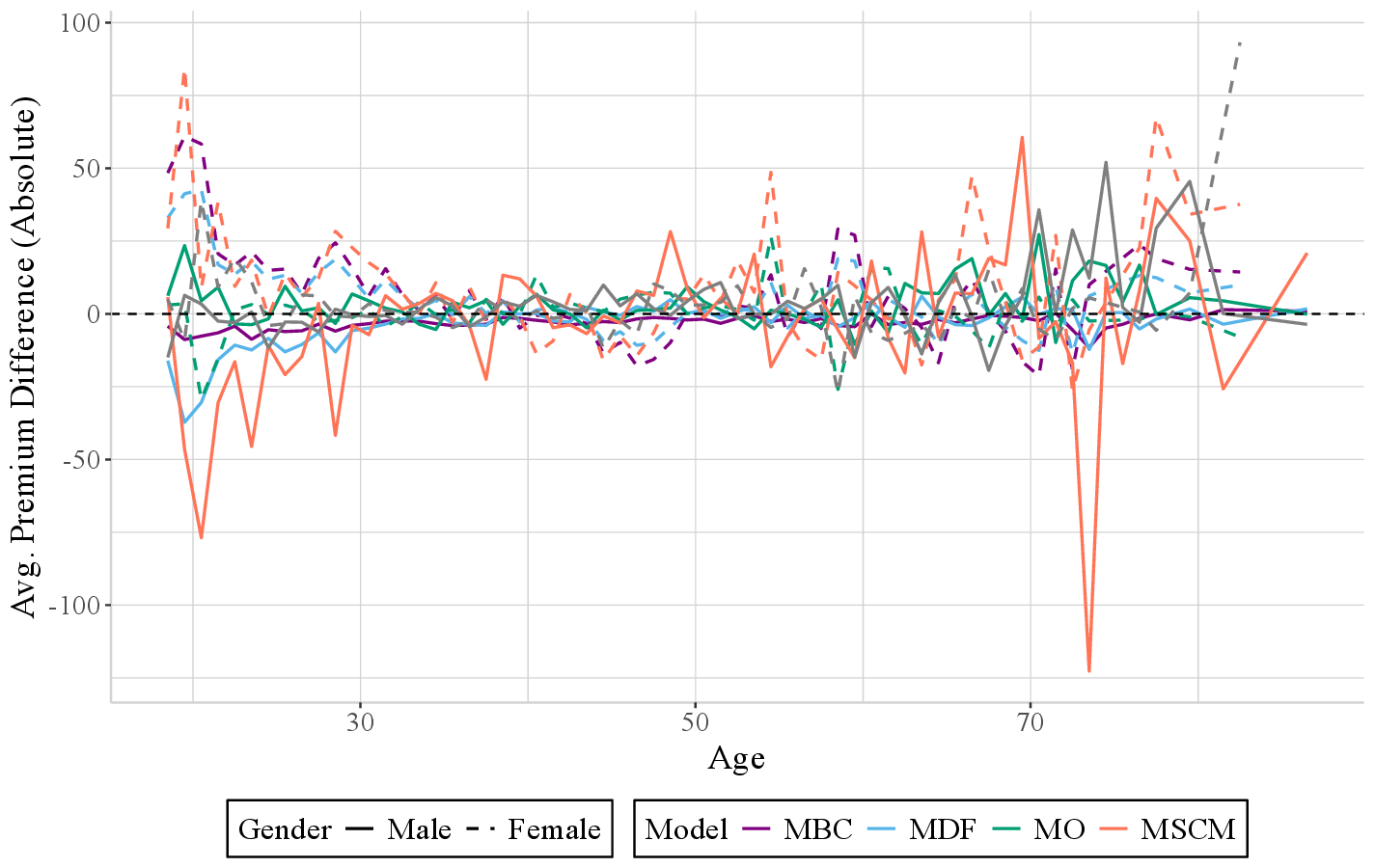}
    \caption{Relative premium difference in \textit{fremotor1prem0304a} (XGBoost models versus XGBoost MB)}
    \label{fig:placeholder2}
\end{figure}

\section{Additional Adverse Selection Plots}\label{as}
In this appendix, we report the remaining double lift charts by gender, supplementing Figures \ref{as fre}-\ref{as pg} in Section \ref{sec:as}.
\begin{figure}[H]
    \centering
    \includegraphics[width=0.65\linewidth]{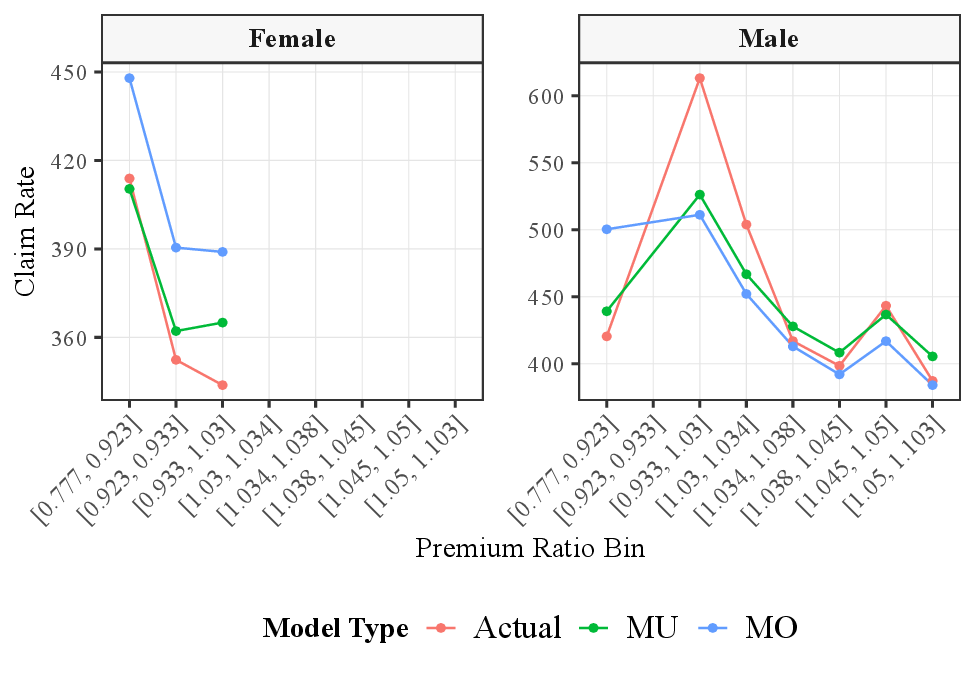}
    \caption{Double lift charts by gender (GLM MO versus GLM MU) in \textit{fremotor1prem0304a}}
    \label{fig:placeholder3}
\end{figure}
\begin{figure}[H]
    \centering
    \includegraphics[width=0.65\linewidth]{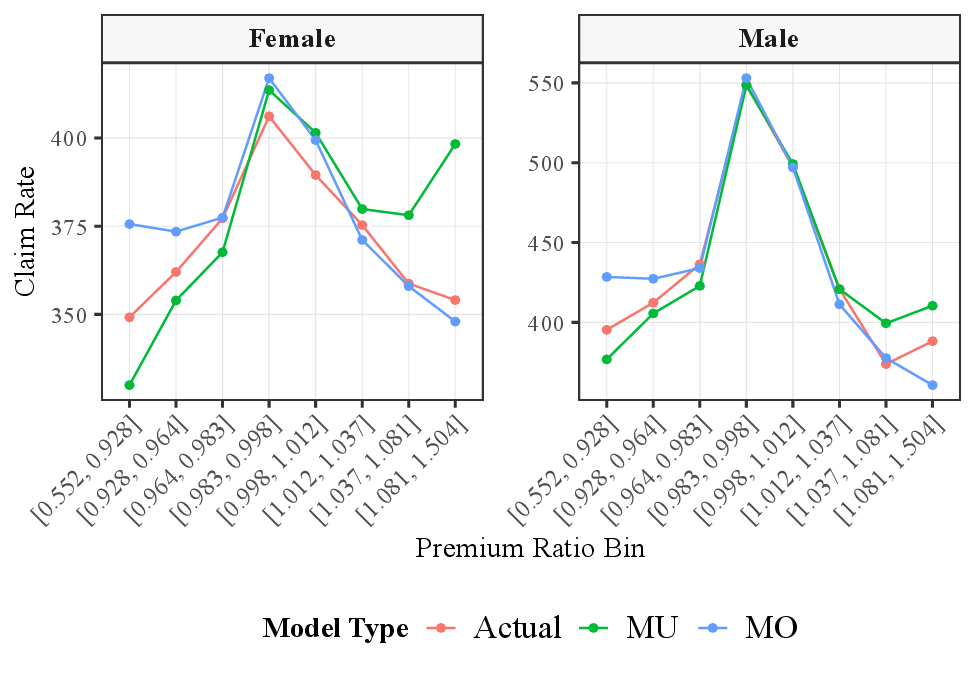}
    \caption{Double lift charts by gender (XGBoost MO versus XGBoost MU) in \textit{fremotor1prem0304a}}
    \label{fig:placeholder4}
\end{figure}

\begin{figure}[H]
    \centering
    \includegraphics[width=0.65\linewidth]{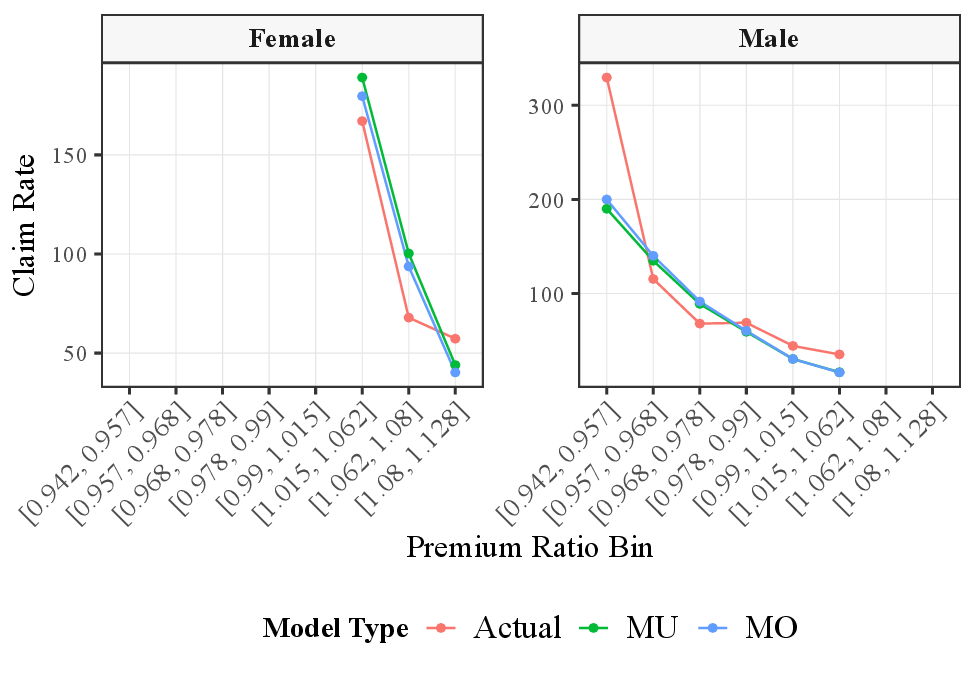}
    \caption{Double lift charts by gender (GLM MO versus GLM MU) in \textit{pg15training}}
    \label{fig:placeholder5}
\end{figure}
\begin{figure}[H]
    \centering
    \includegraphics[width=0.65\linewidth]{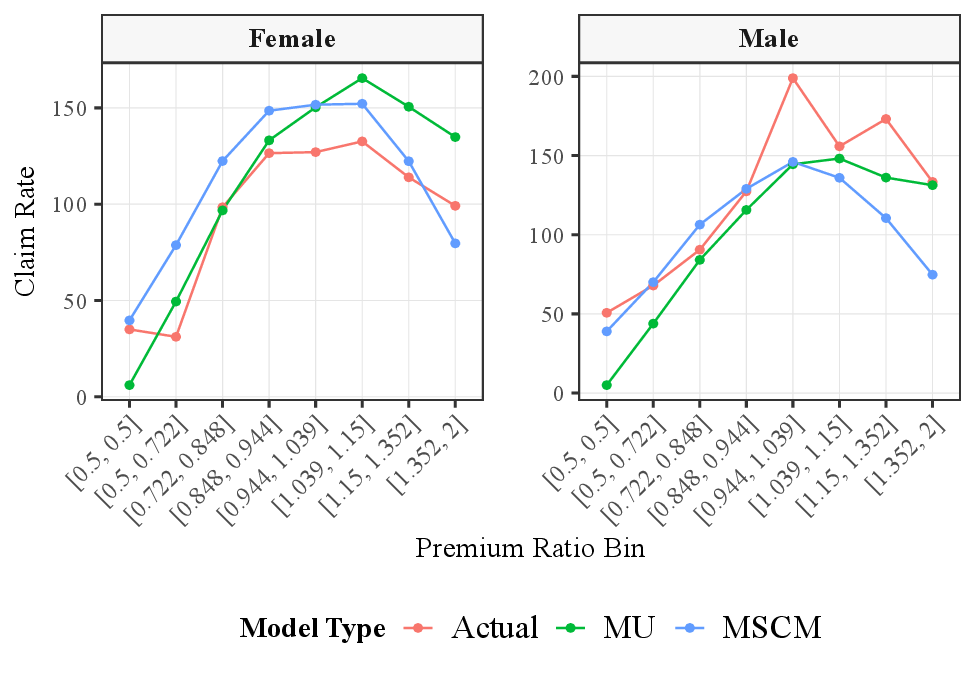}
    \caption{Double lift charts by gender (XGBoost MSCM versus XGBoost MU) in \textit{pg15training}}
    \label{fig:placeholder6}
\end{figure}

\section{Details on NSGA-II Algorithm}\label{nsga}
The NSGA-II algorithm preserves the best solutions across generations, ensuring monotonic improvement and convergence toward the true Pareto front \citep{liu2019improved}. It achieves a computational complexity of $O(MN^2)$ , where $M$ denotes the number of objectives and $N$ the population size, which supports scalability to real-world problems with a moderate number of objectives \citep{verma2021comprehensive}. NSGA-II simultaneously maintains a spread out and well-converged approximation of the Pareto front without requiring prior specification of user preferences. This makes it a computationally efficient, theoretically sound, and empirically validated approach for navigating complex, high-dimensional trade-off landscapes—particularly well-suited for fairness-constrained optimization tasks, where the Pareto frontier is typically unknown, potentially non-convex, and subject to ethical interpretation.

\section{MNN Hyperparameter Tuning}\label{lambda}
To jointly optimize predictive accuracy and counterfactual fairness in the MNN framework, we calibrate the trade-off hyperparameter $\lambda$, which governs the relative weight of the fairness regularization term. Predictive accuracy is quantified by the validation loss, while counterfactual fairness is measured by the average absolute disparity between the observed and counterfactual model outputs: $f^{\text{real}}(x_i) - f^{\text{counterfactual}}(x_i')$.

\begin{figure}[H]
    \centering
    \includegraphics[width=0.85\linewidth]{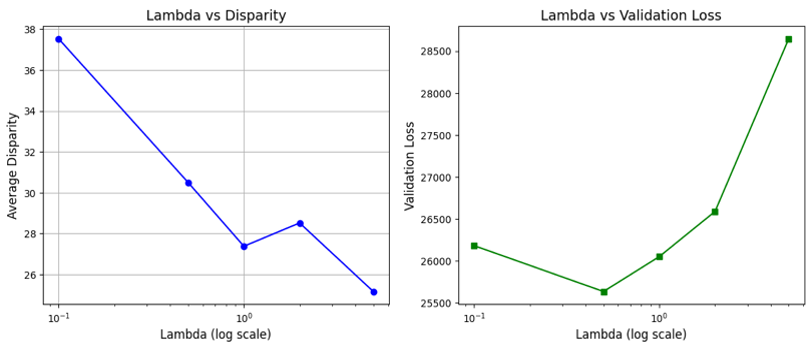}
    \caption{Tuning $\lambda$ in composite loss function in \textit{fremotor1prem0304a}}
    \label{fig:loss-fre}
\end{figure}

\begin{figure}[H]
    \centering
    \includegraphics[width=0.85\linewidth]{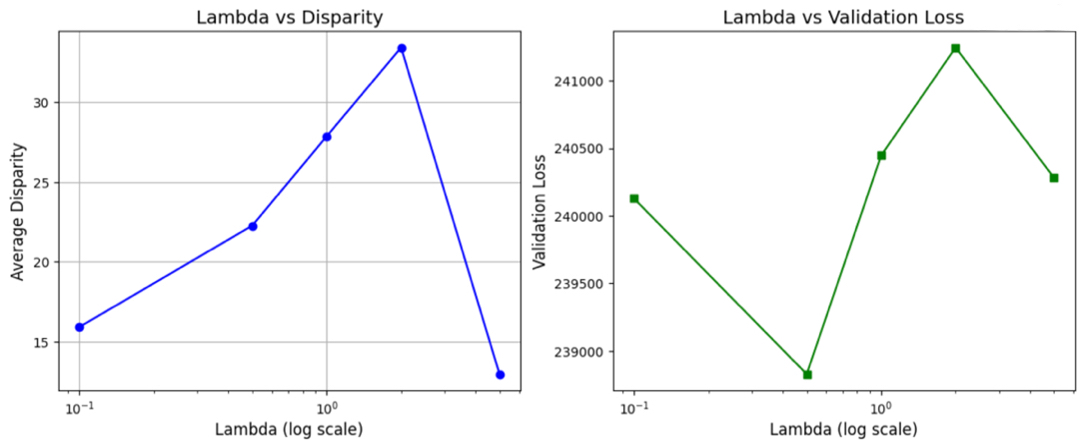}
    \caption{Tuning $\lambda$ in composite loss function in \textit{pg15training}}
    \label{fig:loss-pg}
\end{figure}

We perform stratified 5-fold cross-validation to assess robustness across data partitions. Figures~\ref{fig:loss-fre} and \ref{fig:loss-pg} plot the mean validation loss and mean counterfactual disparity against $\log_{10}(\lambda)$. We select the optimal $\lambda^*$ as a balance point where both validation loss and counterfactual disparity are jointly low.

In dataset \textit{fremotor1prem0304a}, the optimal $\lambda$ is set equal to 1 based on Figure \ref{fig:loss-fre}. 

In dataset \textit{pg15training}, the optimal $\lambda$ is set equal to 5 based on Figure \ref{fig:loss-pg}.

\label{lastpage}
\end{document}